\documentclass[twocolumn,showpacs,preprintnumbers,amsmath,amssymb,pre]{revtex4}

\usepackage{graphicx}
\usepackage{amsfonts}

\begin{document}

\title{Renormalization of Oscillator Lattices with Disorder}

\author{Per \"{O}stborn}
\affiliation{Division of Mathematical Physics, Lund University, S--221 00 Lund, Sweden}

\begin{abstract}
A real-space renormalization transformation is constructed for lattices of non-identical
oscillators with dynamics of the general form
$d\phi_{k}/dt=\omega_{k}+g\sum_{l}f_{lk}(\phi_{l},\phi_{k})$.
The transformation acts on ensembles of such lattices.
Critical properties corresponding to a second order phase transition towards
macroscopic synchronization are deduced. The analysis is potentially exact,
but relies in part on unproven assumptions. Numerically, second order phase transitions
with the predicted properties are observed as $g$ increases in two structurally different,
two-dimensional oscillator models. One model has smooth coupling
$f_{lk}(\phi_{l},\phi_{k})=\varphi(\phi_{l}-\phi_{k})$, where $\varphi(x)$ is non-odd.
The other model is pulse-coupled, with $f_{lk}(\phi_{l},\phi_{k})=\delta(\phi_{l})\varphi(\phi_{k})$.
Lower bounds for the critical
dimensions for different types of coupling are obtained. For non-odd coupling,
macroscopic synchronization cannot be ruled out for any dimension $D\geq 1$,
whereas in the case of odd coupling, the well-known result that it can be ruled out
for $D< 3$ is regained. 
\end{abstract}


\maketitle

\section{introduction}
\label{intro}
The study of synchronization in large oscillator networks has been a thriving field of research
ever since the classic work by Winfree in 1967 \cite{sync1}.
Even so, there are still basic questions that await satisfactory answers. One such
question is when and how macroscopic synchronization occurs in lattices of non-identical oscillators.
This is the subject of the present paper.

Part of the charm of the study of synchronization is that the applications are very diverse
\cite{sync2,sync3,sync4}.
Rhythmic activities in living organisms are, in many cases, generated by the collective
oscillation of a large, synchronized assembly of pacemaker cells. Examples include the beating of the heart \cite{heart}, locomotion \cite{loco},
the circadian rhythm \cite{circadian}, and the peristaltis of the small intestine \cite{intestine}.
There is also growing evidence that large-scale synchronization among
neurons is crucial in the interpretation of sensory data and in conscious perception \cite{perception}.
Epileptic seizures correspond to an abnormal degree of synchronization \cite{epilepsy}.
On a larger scale, synchronization can be seen in groups of organisms.
Swarms of fireflies may flash in unison  \cite{buck}, the chirping of crickets in a field waxes and wanes in
partial synchrony \cite{walker}, an audience may spontaneously start to clap in unison \cite{clapping},
females living together synchronize their menstrual cycles \cite{menstrual}. It has also been realized that
synchronization is an essential concept in the dynamics of spatially extended animal populations \cite{blasius}.
Examples from outside biology include synchronization in power grids \cite{sync4},
among lasers \cite{lasers}, oscillatory chemical reactions \cite{chemical}
and in arrays of Josephson junctions \cite{josephson}.

In most applications, there will inevitably be some variation among the oscillators,
for instance in the \emph{natural frequency} with which they oscillate when isolated.
In this situation, macroscopic synchronization means that the order parameter $r$
becomes non-zero, where
\begin{equation}
r=\lim_{N\rightarrow\infty}M/N.
\label{rdef}
\end{equation}
Here, $M$ is the size of the largest group of oscillators
that attain the same mean frequency and $N$ is the total number of oscillators.
If the network has spatial structure,
the $M$ synchronized oscillators typically form a percolating cluster \cite{saka,ptorus}.
The mean frequency $\Omega_{k}$ of oscillator $k$ is defined as
\begin{equation}
\Omega_{k}=\lim_{t\rightarrow\infty}\phi_{k}(t)/t,
\label{meandef}
\end{equation}
where $\phi_{k}$ is the phase of $k$.
The existence of the above limits has to be assumed \cite{aboutmean}. 

In theoretical work, the description of each oscillator must be simple to enable
the study of large networks. Kuramoto and others \cite{kura} introduced the so
called phase reduction technique, and showed that in the limits of small coupling
between oscillators and small variation of natural frequencies,
the phase $\phi_{k}$ is sufficient to describe the state of
each oscillator $k$, and the network dynamics is given by
\begin{equation}
\frac{d\phi_{k}}{dt}=\omega_{k}+g\sum_{l=1}^{N}\varphi_{lk}(\phi_{l}-\phi_{k}).
\label{phasered}
\end{equation}
The constant $\omega_{k}$ is the natural frequency, $g$ is the coupling strength,
and $\varphi_{lk}(x)$ is a 1-periodic function.
Another situation where a phase description is sufficient is in the limits of short, pulse-like
interactions and strong dissipation. The quick reduction of phase space volumes
then ensures that after one perturbation from a nearby oscillator $l$,
oscillator $k$ returns close to its limit cycle before the next perturbation, and we may write
\begin{equation}
\frac{d\phi_{k}}{dt}=\omega_{k}+g\sum_{l=1}^{N}\delta[\mathrm{mod}(\phi_{l},1)]\varphi_{lk}(\phi_{k}),
\label{pulsemodel}
\end{equation}
where $\delta(x)$ is the Dirac delta function.
This is often a good description in biological applications, where the
interactions, for instance, may consist of electric discharges or light flashes.
If we define $\phi_{k}$ as a cyclic variable, $\phi_{k}\in[0,1)$, we can replace
$\delta[\mathrm{mod}(\phi_{l},1)]$ with $\delta(\phi_{l})$.
The $1$-periodic function $\varphi_{lk}(x)$ is often called the \emph{phase
response curve}.

Transitions to macroscopic synchronization are
similar to phase transitions in equilibrium systems.
The two main methods to analyze phase transitions are
to make a mean-field description or to use a renormalization group.
So far, most attempts to gain understanding of macroscopic synchronization among non-identical
oscillators have assumed that the oscillators are coupled all-to-all.
This is the mean-field description.
To use a renormalization group, on the other hand, is the natural
way to gain understanding of phase transitions in lattices.

This is the method used in this study. A real-space renormalization scheme is developed
that is potentially exact. It is tested numerically on two structurally different models.
At its present stage of development, however, the scheme has to be called heuristic
since it relies in part on unproven assumptions.

Before describing the approach, let me review very briefly the current state
of knowledge about transitions to macroscopic synchronization in mean-field and lattice models.

\section{Review of related work}

\subsection{Mean-field models}

As a special case of Eq. (\ref{phasered}), Kuramoto \cite{kura} introduced the mean-field model
\begin{equation}
\frac{d\phi_{k}}{dt}=\omega_{k}+\frac{g}{N}\sum_{l=1}^{N}\sin[2\pi(\phi_{l}-\phi_{k})].
\label{kuramodel}
\end{equation}
because of its analytical tractability. Instead of $r$, Kuramoto studied the order parameter
\begin{equation}
R=\lim_{t\rightarrow\infty}\lim_{N\rightarrow\infty}\left|\sum_{k=1}^{N}e^{2\pi i\phi_{k}(t)}\right|,
\end{equation}
and found that there is a critical coupling strength $g_{c}$ such that
\begin{equation}\begin{array}{ll}
R=0 & g<g_{c}\\
R\propto(g-g_{c})^{1/2} & g\geq g_{c}
\end{array}.\label{kurares}\end{equation}
If each $\omega_{k}$ is
chosen independently from a density function $\mathcal{D}_{\omega}$ that is
unimodal and symmetric about its mean $\mu$, the critical coupling is given by
$g_{c}=2/\pi \mathcal{D}_{\omega}(\mu)$.

Since the original
work by Kuramoto, the analysis of model (\ref{kuramodel})
has been refined \cite{reviews}. Also, it turns out that
the exponent $1/2$ in Eq. (\ref{kurares}) changes to $1$ as soon as
non-odd harmonics are added to the coupling function
$\varphi_{lk}(x)=\sin(2\pi x)$ \cite{nonodd}.

Note that the order parameter $R$ measures the degree of \emph{phase synchronization},
whereas $r$ [Eq. (\ref{rdef})] measures the degree of \emph{frequency synchronization}.
A non-zero $R$ implies a non-zero $r$, but the opposite is not true.
In a mean-field model,
$r$ and $R$ typically becomes non-zero at the same critical coupling $g_{c}$.
In a lattice model, waves in the phase field may be expected \cite{sync1,sync2,sync3,sync4, kura,strang,kheowan}
even if the frequencies are synchronized, so that $r>0$ but $R=0$.

Ariaratnam and Strogatz \cite{aria} studied Winfree's original model\begin{equation}
\frac{d\phi_{k}}{dt}=\omega_{k}+\frac{g}{N}\sum_{l=1}^{N}\theta(\phi_{l})\varphi(\phi_{k})
\label{winfree}
\end{equation}
in the special case $\theta(x)=1+\cos(2\pi x)$ and $\varphi(x)=-\sin(2\pi x)$. This model
is similar to the model (\ref{pulsemodel}), with the smooth $1$-periodic influence function $\theta(x)$
replacing the delta pulse. The authors were able to
obtain the phase diagram in the plane spanned by $g$ and $\gamma$,
where $\mathcal{D}_{\omega}$ is uniform with support $[1-\gamma,1+\gamma]$.
Apart from the phases with $r=R=0$ and $r=R=1$, there is a phase of partial
synchrony with $r<1$ and $R<1$, and also phases with partial or complete oscillator death
($d\phi_{k}/dt=0$).

Tsubo et al. \cite{tsubo} studied a similar model, but let the disorder reside in the
phase response curves $\varphi_{lk}(x)$, whereas the natural frequencies were identical.
With $\varphi_{lk}(x)=\cos(\pi a_{k})-\cos(2\pi x - \pi a_{k})$, where $a_{k}$
is a random number from a uniform distribution with support $[a_{\min},a_{\max}]$,
they found a discontinuous transition to macroscopic
synchronization in the phase plane spanned by $a_{\min}$ and $a_{\max}$, in contrast
to the continuous transition in the Kuramoto model, as expressed in Eq. (\ref{kurares}).

\subsection{Lattice models}
\label{latticeintro}

The analysis of oscillator lattices is harder than that of mean-field models,
and less progress has been made.
For cubic lattices with dimension $D$ and dynamics of form (\ref{phasered}) with
$\varphi_{lk}(x)=\varphi(x)$ and odd coupling, $\varphi(-x)\equiv-\varphi(x)$, Daido ruled out states with $r>0$ for
$D\leq 2$ \cite{daido}.
Daido obtained this result using renormalization-like arguments.
With similar methods, Strogatz and Mirollo \cite{strogatz} were able to prove that whenever
$\mathcal{D}_{\omega}$ has non-zero variance, states
with $r=1$ are ruled out for \emph{any} finite $D$. In addition, states with
$0<r<1$ cannot have synchronized clusters which contain macroscopic cubes
(with volume $V=aN$, $0<a<1$). Thus, for odd coupling, macroscopic synchronization
may occur only if $D\geq 3$ and can only be partial, with sponge-like synchronized
clusters. Whether such states actually exist is still an open question. The numerical
evidence is inconclusive in my view \cite{saka,daido,aoyagi,hong}.

Kopell and Ermentrout were the first to point out that non-odd coupling
facilitates synchronization \cite{kopell}.
For an oscillator chain ($D=1$) of form (\ref{phasered}) with $\varphi_{lk}(x)=\varphi(x)$,
I studied the case when $\mathcal{D}_{\omega}$ has finite support
$[\omega_{\min},\omega_{\max}]$. For models with $\varphi(0)=0$ and $\varphi'(0)>0$,
there is then a critical
coupling $g_{c}$ at which a discontinuous transition from $r=0$ to $r=1$ takes place \cite{chain1}.
I found that $g_{c}=(\omega_{\max}-\omega_{\min})/\left|d(\hat{x})\right|$,
where the denominator $\left|d(\hat{x})\right|$ is a mesure of the "non-oddity" of $\varphi(x)$,
vanishing for odd coupling such as $\varphi(x)=\sin(2\pi x)$. A similar result was provided for a
model of the form (\ref{pulsemodel}), which can be seen as
inherently non-odd due to the sequential interaction of two oscillators via pulses \cite{chain2}.

Since macroscopic synchronization is possible for $D=1$ for non-odd coupling,
it is expected to be possible for all $D>1$. However, no proofs have been obtained,
to my best knowledge. For $D=2$, me and my co-workers \cite{ptorus} offered numerical evidence for
a continuous, second order phase transition to $r>0$ in a model of form (\ref{pulsemodel}).

In equilibrium systems, there is typically an upper critical dimension,
above which a lattice model shows mean field critical behavior. Hong, Park
and co-workers \cite{hong} have re-examined the lattice version of the Kuramoto model
(\ref{kuramodel}), and claim that $D=4$ is the
upper critical dimension, above which critical exponents take mean field values
and macroscopic frequency- and phase synchronization appear at the same critical
coupling. However, the results by Strogatz and Mirollo \cite{strogatz} indicate that
the upper critical dimension is infinity for this model, since they ruled out states
with $r=1$ for any finite $D$ and any $\mathcal{D}_{\omega}$ with non-zero variance,
whereas such states exist in the mean-field model (\ref{kuramodel}) when $\mathcal{D}_{\omega}$
has non-zero variance, but finite support. It is conceivable that the phase transition
structure of oscillator networks is richer than in equilibrium systems,
and cannot be fully captured by the concepts used there.

Lattice models of oscillator networks are closely related to spatial continuum models.
This is the natural way to describe the oscillatory Belousov-Zhabotinsky reaction \cite{belousov}
and the smooth muscle tissue in the intestine \cite{intestine}.
It may also be an adequate model of a large piece of oscillatory cardiac muscle,
even though it consists of discrete cells.
The preferable mathematical description is given by the Ginzburg-Landau equation (GLE) \cite{sync3,kura},
where the state at each point in space is given by a complex number, encoding both the phase and
amplitude of oscillation. The GLE corresponds to a lattice of \emph{identical} oscillators.
Using a field theoretic renormalization group, Risler and co-workers have performed a thorough
analysis of synhronization transitions in the GLE with noise \cite{risler}. The noise is assumed to be uncorrelated
in space and time. In contrast, random natural frequencies correspond to "noise" that is uncorrelated in space,
but quenched in time. This makes the problem much more difficult in the continuum formulation.
In particular, discontinuities arise in the phase field whenever frequency synchronization
is not perfect ($r<1$).

\section{Models and methods}

\begin{figure}
\begin{center}
\includegraphics[clip=true]{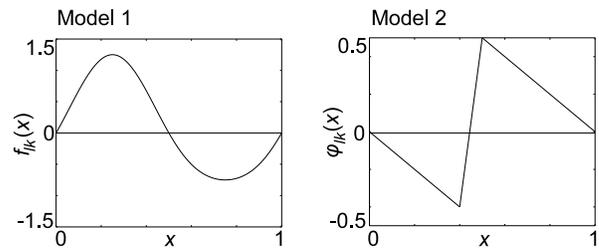}
\end{center}
\caption{Coupling functions in the two test models. Model 1 is of form (\ref{phasered})
with $\varphi_{lk}(x)$ given by Eq. (\ref{model1}).
Model 2 has form (\ref{pulsemodel}) with $\varphi_{lk}(x)$ given by Eq. (\ref{model2})}
\label{Rfig1}
\end{figure}

In the analysis, models of the following form are considered:
\begin{equation}
\frac{d\phi_{k}}{dt}=\omega_{k}+g\sum_{l\in n_{k}}f_{lk}(\phi_{l},\phi_{k}),\ \ \ \ \ \ k=1,\ldots, N.
\label{model}
\end{equation}
Here, $\phi_{k}\in \mathbf{\mathbb{R}}$ is the phase of oscillator $k$, $\omega_{k}$ is its
natural frequency, $g$ is the coupling strength, and  $n_{k}$ is the set of $k$'s nearest neighbors.
The analysis is restricted to cubic lattices of dimension $D$.
The coupling functions $f_{lk}$ are assumed to be $1$-periodic in each argument.
With this restriction, the phases
$\phi_{k}$ are allowed to grow linearly to be able to count the number of cycles,
that is, the largest integer smaller than $\phi_{k}(t)-\phi_{k}(0)$.
Since no further assumptions are made, the results are expected to apply (at least) to all models
of this form. All the coupling functions in the models referred to above have the form
given in Eq. (\ref{model}).

Let us define the ensemble
\begin{equation}
\mathcal{E}=\mathcal{E}(g,\mathcal{D}_{\omega},\mathcal{D}_{f},\mathcal{D}_{\phi(0)},D,N)
\label{ensemble}
\end{equation}
of realizations of systems (\ref{model}), where $\omega_{k}$ are independent random numbers
from the density function $\mathcal{D}_{\omega}$,
each $f_{lk}$ is chosen from $\mathcal{D}_{f}$, and the initial condition
$\phi(0)=[\phi_{1}(0),\ldots,\phi_{N}(0)]$ is chosen from $\mathcal{D}_{\phi(0)}$.
Quenched disorder is introduced by $\mathcal{D}_{\omega}$ and $\mathcal{D}_{f}$.

To give the coupling strength $g$ a clear meaning, $\mathcal{D}_{f}$ should be chosen so that
\begin{equation}
\langle\int_{0}^{1}\int_{0}^{1}|f_{lk}(\phi_{l},\phi_{k})|d\phi_{l}d\phi_{k}\rangle_{\mathcal{D}_{f}}=1,
\end{equation}
or so that it fulfils a similar condition. Alternatively, one may drop $g$ as an argument
of $\mathcal{E}$.

To test the theoretical predictions, numerical simulations
of two specific models with $D=2$ are performed.
The first model has the form (\ref{phasered}) with
\begin{equation}
\varphi_{lk}(x)=\sin(2\pi x)+\frac{1}{4}\sin^{2}(2\pi x),
\label{model1}\end{equation}
(Fig. \ref{Rfig1}). This model will be referred to as \emph{Model 1}.
The density function $\mathcal{D}_{\omega}$ is uniform with support $[1.0,1.5]$
and $\mathcal{D}_{\phi(0)}$ is uniform in the interval $[0,1]$.
Forward Euler integration is used, with $\Delta t = 0.05$.
The motivation for this model is that it is similar to the Kuramoto model,
but has a non-odd term to allow macroscopic synchronization for $D=2$ (see below).

The second model has the form (\ref{pulsemodel}) with
\begin{equation}
\varphi_{lk}(x)=
\left\{\begin{array}{ll} -x, & x\leq 0.4\\
                         9x-4, & 0.4<x<0.5\\
                         1-x, & 0.5\leq x<1
\end{array}\right.,
\label{model2}\end{equation}
(Fig. \ref{Rfig1}). This model will be referred to as \emph{Model 2}.
$\mathcal{D}_{\omega^{-1}}$ is uniform with support $[1.0,1.5]$
and $\mathcal{D}_{\phi(0)}$ is uniform in the interval $[0,1]$.
The same integration method as in Refs. \cite{ptorus} and \cite{chain1} is used.
The piecewise linear phase response curve expressed by Eq. (\ref{model2})
has the same bipolar characteristics
as the curve $\varphi(x)=-\sin(2\pi x)$ used by Ariaratnam and Strogatz \cite{aria}.
This type of response to external perturbation is found in many biological applications
\cite{sync2}.

The details are arbitrary, but these models are chosen since they have been studied previously,
model (\ref{model1}) for $D=1$ in Ref. \cite{chain1} and model (\ref{model2})
for $D=1$ in Ref. \cite{chain2} and in the case $D=2$ in Refs. \cite{ptorus} and \cite{strang}.

Unless otherwise stated, simulations of Model 1 are carried out with lattice size $300\times 300$,
whereas for Model 2 the lattice size $500\times 500$ is used.
The larger lattice size is needed to resolve phase 2 in Model 2 (see below).
Periodic boundary conditions are used, unless otherwise stated. 
In and around critical regions, a transient time of $t= 100 000 - 200 000$ is used
before measurements are done. For $g$ well below critical values, shorter transient times are used.
Mean frequencies $\Omega_{k}$ are approximated by taking the mean of $d\phi_{k}/dt$
during a time interval $\Delta t=1000$ after the initial transient \cite{aboutmean}.

To identify frequency clusters, the lattice is scanned. An oscillator $k$ is considered
to belong to the same cluster as a previously scanned neighbor oscillator $l$ if
$|\Omega_{l}-\Omega_{k}|<0.001$. If two such neighbor oscillators $l$ and $l'$ are preliminarily
judged to belong to different clusters $C$ and $C'$, but both fulfil the above inequality,
$C$ and $C'$ are identified as two parts of the same cluster.

\section{Theoretical approach}

\begin{figure}
\begin{center}
\includegraphics[clip=true]{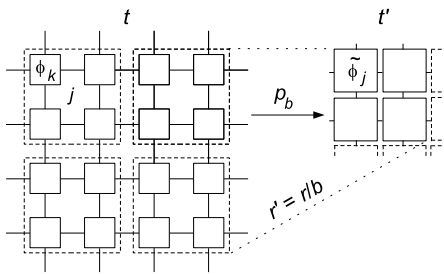}
\end{center}
\caption{A block-oscillator transformation (\ref{blocktrans}) with scale factor $b=2$
         and implicit change of length scale $r'=r/b$.}
\label{Rfig2}
\end{figure}

The first step in the renormalization scheme is to define a block-oscillator
transformation $p_{b}: \mathbf{\mathbb{R}}^{b^{d}+1}\rightarrow\mathbf{\mathbb{R}}^{2}$ (Fig. \ref{Rfig2})
\begin{equation}
[\tilde{\phi}_{j}(t'),t']=p_{b}[\{\phi_{k}(t)\}_{k\in j},t].
\label{blocktrans}
\end{equation}
We get a coarse-grained version of the lattice and interpret the phase $\tilde{\phi}_{j}$
as the state of block oscillator $j$. Applying $p_{b}$ to all $j$, we may define
the scale transformation $P_{b}: \mathbf{\mathbb{R}}^{N+1}\rightarrow\mathbf{\mathbb{R}}^{N/b^{D}+1}$ as
\begin{equation}
[\tilde{\phi}(t'),t']=P_{b}[\phi(t),t],
\end{equation}
with $\phi=(\phi_{1},\ldots,\phi_{N})$
and $\tilde{\phi}=(\tilde{\phi}_{1},\ldots,\tilde{\phi}_{N/b^{D}})$.

\begin{figure}
\begin{center}
\includegraphics[clip=true]{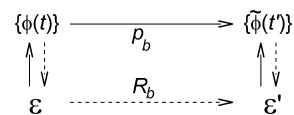}
\end{center}
\caption{Relation between the renormalization transformation $R_{b}$ and the block-oscillator
transformation $p_{b}$. An ensemble $\mathcal{E}$ produces an infinite set $\{\phi(t)\}$ of
time series $\phi(t)$. Only if such a set $\{\phi(t)\}$ inversely determines $\mathcal{E}$
(dashed vertical arrows), is $\mathcal{E}'$ uniquely determined by $\{\tilde{\phi}(t')\}$.
Thus $R_{b}$ does only exist in this case.}
\label{Rfig3}
\end{figure}

Let us discuss these transformations in a bit more detail.
To be able to interpret $P_{b}$ as a scale transformation, note that it must fulfil the
group property
\begin{equation}
P_{b_{1}}P_{b_{2}}=P_{b_{1}b_{2}}.
\label{group}
\end{equation}
Regarding $p_{b}$, I restrict the interst to linear transformations of the form
\begin{equation}
\left(\begin{array}{c}
\tilde{\phi}_{j}\\t'\end{array}\right)=M(b,D)
\left(\begin{array}{c}
\phi_{k_{1}}\\\vdots\\\phi_{k_{b^{D}}}\\t\end{array}\right),
\label{transform}
\end{equation}
where $\phi_{k_{1}},\ldots,\phi_{k_{b^{D}}}$ is some list of the phases in block $j$,
and $M(b,D)$ is a $2\times(b^{D}+1)$-matrix with
\[
M(b,D)=\left(\begin{array}{cccc}
A & \ldots & A & B\\
0 & \ldots & 0 & C
\end{array}\right).
\]
(Here, $A$, $B$, and $C$ are functions of $b$ and $D$.) Linear transformations
with $M_{1,1} = \ldots = M_{1,b^{D}}=A$ are considered since $\tilde{\phi}_{j}$
is intended to be a kind of arithmetic mean of the phases $\phi_{k}$ in block $j$,
where all $\phi_{k}$ are treated in the same way. Such a choice of $\tilde{\phi}_{j}$
is justified if the phases are interpreted to be linear variables, instead of cyclic ones.
There is no reason to let the transformed time $t'$ depend on the phases,
and therefore I set $M_{2,1} = \ldots = M_{2,b^{D}}=0$.

Next, let $\left\langle H[\phi(t)]\right\rangle_{\mathcal{E}}$
be the ensemble mean of $H[\phi(t)]$, where $H$ is a functional of the
time series $\phi(t)$. We then seek an ensemble
\begin{equation}
\mathcal{E}'=\mathcal{E}(g',\mathcal{D}_{f'},\mathcal{D}_{\omega'},\mathcal{D}_{\phi'(0)},D,N/b^{D})
\label{ensemble2}
\end{equation}
[c.f. Eq. (\ref{ensemble})] such that
\begin{equation}
\langle H[\phi'(t')]\rangle_{\mathcal{E}'}=\langle H[\tilde{\phi}(t')]\rangle_{\mathcal{E}}
\label{ensemble2condition}
\end{equation}
for any functional $H$. In words, all statistical quantities produced by the desired
ensemble $\mathcal{E}'$ should be the same as those given by the transformed
phases $\{\tilde{\phi}(t')\}$, which in turn are determined by $\mathcal{E}$. The relation
\begin{equation}
\mathcal{E}'=R_{b}\mathcal{E}
\end{equation}
defines the renormalization transformation $R_{b}$,
assuming that $\mathcal{E}'$ exists (Fig. \ref{Rfig3}).

Naively, instead of working at the ensemble level, we could have looked for an evolution equation
for the transformed phases $\tilde{\phi}$. If we would have been successful,
we could have written
$d\tilde{\phi}_{j}/dt'=\omega_{j}'+g'\sum_{i\in n_{j}}f_{ij}'(\tilde{\phi}_{i},\tilde{\phi}_{j})$ or,
in compressed form, $d\tilde{\phi}_{j}/dt'=\omega_{j}'+h'(\tilde{\phi})$. However, since
$p_{b}$ is not invertible, we have to express the interaction in the original phases,
i.e. $h'=h'(\phi)$, and we do not get an evolution equation of the transformed phases in closed form
such as Eq. (\ref{model}).

Let us write
\begin{equation}\begin{array}{lll}
d\phi_{k}/dt & = & \omega_{k}+h_{k}(\phi)\\
d\phi_{k}'/dt' & = & \omega_{k}'+h_{k}'(\phi')
\end{array}
\label{shorthand}\end{equation}
for the original and transformed ensembles $\mathcal{E}$ and $\mathcal{E}'$, respectively.
In the following, I also adopt the notation $\mathrm{E}[x]=\langle x\rangle_{\mathcal{E}}$ and
let $\mathrm{Var}[x]$ and $\mathrm{Cov}[x,y]$ be the ensemble variance and
covariance, respectively.

\begin{figure}
\begin{center}
\includegraphics[clip=true]{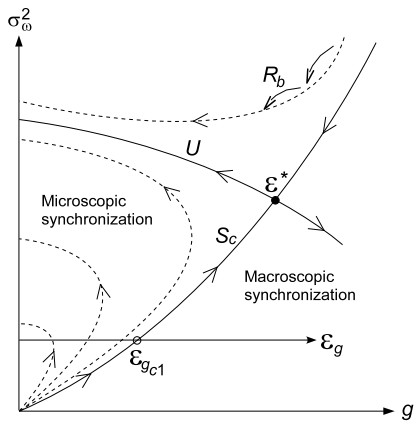}
\end{center}
\caption{Projection of $\{\mathcal{E}\}$
         to the plane spanned by coupling strengh $g$ and the variance $\sigma^{2}_{\omega}$ of natural frequencies.
         The transformation $p_{b}$ [Eq. (\ref{blocktrans})] is chosen so that a critical fixed point $\mathcal{E}^{*}$
         with finite $g^{*}$ and $(\sigma_{\omega}^{2})^{*}$ may appear. The flow under $R_{b}$ is intended
         to be such that ensembles on the critical line $S_{c}$ are attracted to $\mathcal{E}^{*}$, and that
         $\sigma^{2}_{\omega}$ is invariant at $g=0$. 
         Numerically, two ensemble families $\mathcal{E}_{g}$ are studied, Model 1 and Model 2.
         Each of these seemingly becomes critical at some coupling $g=g_{c1}$.}
\label{Rfig4}
\end{figure}

To be able to extract information about critical behavior from $R_{b}$, the transformation
$p_{b}$ has to be chosen so that there may appear a non-trivial fixed point ensemble
\begin{equation}
\mathcal{E}^{*}=R_{b}\mathcal{E}^{*}
\label{fixedpoint}
\end{equation}
in the limit $N\rightarrow\infty$ (Fig. \ref{Rfig4}).
In this study I look for, and assume the existence of,
a fixed point $\mathcal{E}^{*}$ for which the variance of natural frequencies and the variance
of the interaction exist, that is
\begin{equation}\begin{array}{c}
0<(\sigma_{\omega}^{2})^{*}<\infty\\
0<\mathrm{Var}^{*}[h_{k}]<\infty
\end{array}.
\label{finitevar}
\end{equation}
A finite $\mathrm{Var}^{*}[h_{k}]$ implies a finite fixed point coupling strength $g^{*}$ \cite{finiteg}.

Further, $\mathcal{E}^{*}$ should attract an ensemble $\mathcal{E}_{g_{c1}}$ belonging to a family
$\mathcal{E}_{g}$ that passes a transition to macroscopic synchronization at the critical coupling $g=g_{c1}$.
The behavior of $\mathcal{E}_{g_{c}}$ will then be the same as that of $\mathcal{E}^{*}$
at large scales and after long times.

With this in mind, $p_{b}$ is chosen to fulfil three conditions,
in addition to Eqs. (\ref{group}) and (\ref{transform}). Before stating
these conditions, let me introduce a few quantities.

First, let $m(t)$ and $m_{\infty}$ be mean attained frequencies:
\begin{equation}\begin{array}{lll}
m(t) & = & \langle d\phi_{k}/dt\rangle_{\mathcal{E}}\\
m_{\infty} & = & \lim_{t\rightarrow\infty}m(t).
\end{array}\end{equation}
The limit $\lim_{t\rightarrow\infty}m(t)$ exists at the presumed fixed point $\mathcal{E}^{*}$
according to assumption (\ref{finitevar}).
In fact, it follows from Eqs. (\ref{shorthand}) and (\ref{finitevar}) that
the two first moments of the distribution of attained mean frequencies $\Omega_{k}$
exist at $\mathcal{E}^{*}$ \cite{aboutmean,momentsexist}:
\begin{equation}\begin{array}{c}
E^{*}[\Omega_{k}]=(m_{\infty})^{*}<\infty\\
0<\mathrm{Var}^{*}[\Omega_{k}]<\infty.
\label{varexists}
\end{array}\end{equation}

Further, let $\kappa$ be the mean wave number. Since the phases are allowed to be linear variables,
the wave nature of the phase landscape in a given lattice $\mathcal{L}$ may only be manifest
if a suitable integer $q_{k}$ is added or subtracted
to each $\phi_{k}$. Writing $Q=(q_{1},\ldots,q_{N})$ and $\phi^{(Q)}=\phi+Q$, I define
\begin{equation}
\kappa=\lim_{t\rightarrow\infty}\mathrm{Min}\left[\langle |\phi_{k}^{(Q)}-\phi_{l\in n_{k}}^{(Q)}|\rangle_{\mathcal{L}}\right]_{Q}.
\label{kappadef}
\end{equation}
In other words, $Q$ should be chosen so that the mean phase difference
between neighbor oscillators is minimized, and this phase difference is $\kappa$.
The lattice mean $\langle\ldots\rangle_{\mathcal{L}}$
is expected to become equivalent to an ensemble mean $\langle\ldots\rangle_{\mathcal{E}}$
in the limit $N\rightarrow\infty$. (Otherwise an ensemble mean can be added in the
definition.)

\begin{figure}
\begin{center}
\includegraphics[clip=true]{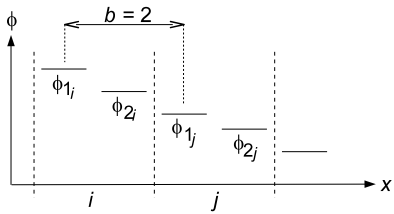}
\end{center}
\caption{Illustration for the argument why condition 3 is fulfilled
for the block-oscillator transformation $p_{b}$ [Eq. (\ref{btrans})]. A wave
is moving in the positive $x$-direction in the case $D=1$. Two neighbor block-oscillators
$i$ and $j$ with size $b=2$ are indicated. Corresponding individual oscillators in the two blocks
are paired as $(\phi_{1_{i}},\phi_{1_{j}})$ and $(\phi_{2_{i}},\phi_{2_{j}})$.
The oscillators in each pair are placed at distance $b=2$ from each other.}
\label{Rfig5}
\end{figure}

Let $\tilde{m}$ and $\tilde{\kappa}$ be the corresponding mean frequency and wave number in the
transformed lattice. The three conditions that guide the choice of $p_{b}$,
apart from Eqs. (\ref{group}) and (\ref{transform}), are then:
\begin{enumerate}
	\item There is a (non-empty) set of ensembles $\Sigma_{1}$, such that if $\mathcal{E}\in \Sigma_{1}$,
	then $\tilde{m}_{\infty}=m_{\infty}$ for any $b$.
	\item There is a set of ensembles $\Sigma_{2}\subseteq \Sigma_{1}$,
	such that if $\mathcal{E}\in \Sigma_{2}$, then $\tilde{\kappa}=\kappa$ for any $b$.
	\item There is a set of ensembles $\mathcal{E}$ with no coupling ($g=0$) such that if $\sigma_{\omega}^{2}$
	is finite and non-zero, then $\sigma_{\omega'}^{2}=\sigma_{\tilde{\omega}}^{2}$ is finite
	and non-zero for any $b$.
\end{enumerate}
It is clear that the two first conditions are necessary.
The critical fixed point $\mathcal{E}^{*}$ I hope to construct belongs to $\Sigma_{2}$.

To see why condition 3 is necessary, look at Fig. \ref{Rfig4}. Assume first that
the condition is broken and that $\lim_{b\rightarrow\infty}\sigma_{\omega'}^{2}=0$
for all ensembles with no coupling.
Then the unstable manifold $U$ of $\mathcal{E}^{*}$ bends down to the origin.
If $S_{c}$ would still be a stable manifold of $\mathcal{E}^{*}$, closed flow lines
would appear, which is impossible since correlation
lengths are always reduced a factor $b$ each time $R_{b}$ is applied.
Thus the flow along the critical line $S_{c}$ changes direction, and the critical properties
at $g_{c1}$ are no longer given by those of $\mathcal{E}^{*}$. In other words,
$\mathcal{E}^{*}$ becomes irrelevant.

Assume instead that $\lim_{b\rightarrow\infty}\sigma_{\omega'}^{2}=\infty$ at $g=0$
for all ensembles with no coupling. I will give a plausibility argument why this is
not consistent with the existence of a fixed point $\mathcal{E}^{*}$ of the desired kind.
From Eqs. (\ref{transform}) and (\ref{shorthand}),
$d\tilde{\phi}_{j}/dt'=(A/C)\sum_{k\in j}\omega_{k}+(A/C)\sum_{k\in j}h_{k}+B/C$.
Since $g=0$ corresponds to $h_{k}\equiv 0$, we have
$d\tilde{\phi}_{j}/dt'=\left(\tilde{\omega}_{j}\right)_{g=0}+(A/C)\sum_{k\in j}h_{k}$.
Here, $\left(\tilde{\omega}_{j}\right)_{g=0}$ is the transformed natural frequency of block $j$
in the ensemble obtained when $g$ is replaced by $0$ in the original ensemble $\mathcal{E}(g,\ldots)$.
Taking the ensemble variance and applying the equation to the presumed fixed point
$\mathcal{E}^{*}(g^{*},\ldots)$, we may write $\mathrm{Var}^{*}[d\tilde{\phi}_{j}/dt']=
\left(\sigma_{\omega'}^{2}\right)_{g=0}+R$, where I do not specify the rest term $R$ for clarity.
The left hand side of this equation must be finite for any $b$, since
$\mathrm{Var}[d\tilde{\phi}_{j}/dt']=\mathrm{Var}[d\phi_{k}/dt]$ at a fixed point
and $\mathrm{Var}^{*}[d\phi_{k}/dt]$ exists \cite{momentsexist}.
This condition would be hard to fulfil if $\left(\sigma_{\omega'}^{2}\right)_{g=0}\rightarrow\infty$
as $b\rightarrow\infty$. Then the term $R$ must sensitively balance this divergence.

Straightforward algebra shows that the only block-oscillator transformation $p_{b}$
of form (\ref{transform}) that has the group property (\ref{group}), and satisfies conditions
$1$ - $3$ is the one with
\begin{equation}
\begin{array}{lll}
A & = & b^{-D-1}\\
B & = & (b^{-D/2-1}-b^{-1})m_{\infty}\\
C & = & b^{-D/2-1},
\end{array}\end{equation}
or, explicitly,
\begin{equation}
\left\{
\begin{array}{lll}
\tilde{\phi}_{j}(t') & = & b^{-D-1}\sum_{k\in j}\phi_{k}(t)-b^{-1}(1-b^{-D/2})m_{\infty} t\\
t' & = & b^{-D/2-1} t
\end{array}\right..
\label{btrans}
\end{equation}
(In fact, I only demonstrate that this transformation satisfies condition $2$ in a restricted sense,
to be described below.)
   
Note that if we are interested in critical ensembles for which $\sigma^{2}_{\omega}$ does not exist,
the three conditions, and hence transformation (\ref{btrans}), should be modified. This point is
discussed by Daido \cite{daido}. This in turn affects the
critical properties and the critical dimension. I do not deal with these cases explicitly
in this paper.

Instead of proving the uniqueness of transformation (\ref{btrans}),
let us examine to what extent it fulfils conditions $1$ - $3$.

Regarding condition 1, we get
\begin{equation}
\tilde{m}_{\infty}=m_{\infty}
\label{samemean}
\end{equation}
for any $b$ and any $\mathcal{E}$ by direct evaluation of
$\lim_{t\rightarrow\infty}\langle d\tilde{\phi}_{j}/dt'\rangle_{\mathcal{E}}$,
using Eq. (\ref{btrans}).

Condition 2 is fulfilled in the following restricted sense. If we may choose $Q$ [Eq. (\ref{kappadef})]
such that a plane wave moving along a principal axis appears in the phase field $\phi^{(Q)}(\mathbf{r})$,
then its wave number will reamin the same in the transformed field
$\tilde{\phi}^{(Q)}(\mathbf{r}')$.
Assume that $i$ and $j$ are two neighbor block-oscillators along the relevant principal axis
(Fig. \ref{Rfig5}). Then, dropping the superscript $(Q)$ for brevity,
\begin{equation}
\tilde{\phi}_{j}-\tilde{\phi}_{i}=b^{-D-1}\sum_{k=1}^{b^{D}}(\phi_{k_{j}}-\phi_{k_{i}}),
\label{blockwaves}
\end{equation}
where $k_{i}$ and $k_{j}$ are the oscillators at corresponding positions in block $i$ and $j$,
respectively. The distance between these is $b$, and thus
$\langle\phi_{k_{j}}-\phi_{k_{i}}\rangle_{\mathcal{L}}=b\langle\phi_{l}-\phi_{k}\rangle_{\mathcal{L}}$,
where $k$ and $l$ are neighbor oscillators along the principal axis. Consequently, taking
the lattice mean of Eq. (\ref{blockwaves}), we get 
$\langle\tilde{\phi}_{j}-\tilde{\phi}_{i}\rangle_{\mathcal{L}}=\langle\phi_{l}-\phi_{k}\rangle_{\mathcal{L}}$.

Turning to condition 3, let us use Eqs. (\ref{btrans}) and (\ref{samemean}) to write
\begin{equation}
\frac{d\tilde{\phi}_{j}}{dt'}-\tilde{m}_{\infty}=b^{-D/2}\sum_{k\in j}(\frac{d\phi_{k}}{dt}-m_{\infty}).
\label{fstretch}
\end{equation}
At $g=0$ we have $d\tilde{\phi}_{j}/dt'=\tilde{\omega}_{j}$ and $d\phi_{k}/dt=\omega_{k}$.
Also, $\tilde{m}_{\infty}=m_{\infty}=\tilde{\mu}=\mu$ by Eq. (\ref{samemean}), so that
$\tilde{\omega}_{j}-\tilde{\mu}=b^{-D/2}\sum_{k\in j}(\omega_{k}-\mu)$.
Thus $\sigma_{\tilde{\omega}}^{2}=\sigma_{\omega}^{2}$ for all $b$.
Condition 3 is then fulfilled
since $\mathcal{D}_{\omega'}=\mathcal{D}_{\tilde{\omega}}$ at $g=0$.
In the limit $b\rightarrow\infty$,
$\mathcal{D}_{\omega'}$ becomes a Gaussian by the central limit theorem,
with the same mean and variance as $\mathcal{D}_{\omega}$.

Transformation (\ref{btrans}) is the only block-oscillator transformation of form
(\ref{transform}) that enables a non-trivial fixed point $\mathcal{E}^{*}$ of the desired kind.
There may be acceptable transformations that do not have form (\ref{transform}).
This is not essential. Fixed point properties derived from any $p_{b}$
that gives rise to a fixed point $\mathcal{E}^{*}$ of the desired kind
have to reflect critical properties of model (\ref{model}), if the corresponding family of ensembles
$\mathcal{E}_{g}$ pass through the critical surface $S_{c}$ as coupling strength $g$ is varied
(Fig. \ref{Rfig4}).
 
To gain some information about $R_{b}$, let us try to express
\begin{equation}
d\tilde{\phi}_{j}/dt'=\tilde{\omega}_{j}+\tilde{h}_{j}(\phi),
\label{tildetrans}
\end{equation}
where $\tilde{\omega}_{j}$ is a constant that can be interpreted as the natural
frequency of block-oscillator $j$ in the transformed lattice, and $\tilde{h}_{j}(\phi)$ can
be seen as the interaction term. To allow such an interpretation, $\tilde{h}_{j}(\phi)$
has to be zero when block $j$ is decoupled from the rest of the lattice. Let
\begin{equation}
m_{j}=\lim_{t\rightarrow\infty}b^{-D}\sum_{k\in j}[d\phi_{k}/dt]^{\prec},
\end{equation}
where $x^{\prec}$ is the value of $x$ when block-oscillator $j$ is decoupled from the surroundings.
Further, let an underlined variable denote an intitial condition mean:
\begin{equation}
\underline{x}=\langle x\rangle_{\mathcal{D}_{\phi(0)}}.
\end{equation}
In a sense, $\underline{m}_{j}$ is then the natural frequency of block-oscillator $j$
in the original lattice. Using Eqs. (\ref{model}) and (\ref{btrans}), we may express
\begin{equation}\begin{array}{lll}
\frac{d\tilde{\phi}_{j}}{dt'} & = & m_{\infty}+b^{D/2}(\underline{m}_{j}-m_{\infty})+\\
& & b^{-D/2}\sum_{k\in j}\left[g\left[\sum_{l\in n_{k}}f_{lk}(\phi_{l},\phi_{k})\right]-(\underline{m}_{j}-\omega_{k})\right].
\end{array}\end{equation}
Let us interpret:
\begin{equation}
\tilde{\omega}_{j}=m_{\infty}+b^{D/2}(\underline{m}_{j}-m_{\infty}),
\label{transomega}
\end{equation}
and
\begin{equation}\begin{array}{lll}
\tilde{h}_{j}(\phi) & = & b^{-D/2}\sum_{k\in j}\left\{g\left[\sum_{l\in n_{k}}f_{lk}(\phi_{l},\phi_{k})\right]-(\underline{m}_{j}-\omega_{k})\right\}\\
& = & b^{-D/2}\sum_{k\in j}g\sum_{l\in n_{k}}[f_{lk}(\phi_{l},\phi_{k})-\underline{f}_{lk}^{\prec}]\\
& = & b^{-D/2}\sum_{k\in j}(h_{k}(\phi)-\underline{h}_{k}^{\prec}).
\end{array}\label{transh}\end{equation}
Here, $\underline{f}_{lk}^{\prec}$ is the initial condition mean of the coupling function $f_{lk}(\phi_{l},\phi_{k})$
as $t\rightarrow\infty$ and block $j$ is decoupled, and $\underline{h}_{k}^{\prec}=g\sum_{l\in n_{k}}\underline{f}_{lk}^{\prec}$.
The interaction $\tilde{h}_{j}$ in Eq. (\ref{transh}) is not strictly zero when $j$ is decoupled
from other blocks. However, the initial condition mean $\underline{\tilde{h}}_{j}$ \emph{is} zero,
in the limit $t\rightarrow\infty$. Thus, these identifications
can be used to deduce \emph{asymptotic} critical behavior of \emph{initial condition averaged} variables,
but nothing else.

I formulate the following conjecture:
\begin{equation}\begin{array}{lll}
\mathcal{D}_{\omega'} & = & \mathcal{D}_{\tilde{\omega}}\\
\langle H[\underline{h}_{k}'(\phi')]\rangle_{\mathcal{E}'} & = & \langle H[\underline{\tilde{h}}_{j}(\phi)]\rangle_{\mathcal{E}}
\end{array}
\label{conjecture}
\end{equation}
for any functional $H$ as $t\rightarrow\infty$.
A number of critical properties follow from Eqs. (\ref{ensemble2condition}),
(\ref{btrans}), (\ref{conjecture}), and the fixed point condition
$\mathcal{E}'=\mathcal{E}=\mathcal{E}^{*}$.

\section{Results}

\subsection{Phase diagrams}

\begin{figure}
\begin{center}
\includegraphics[clip=true]{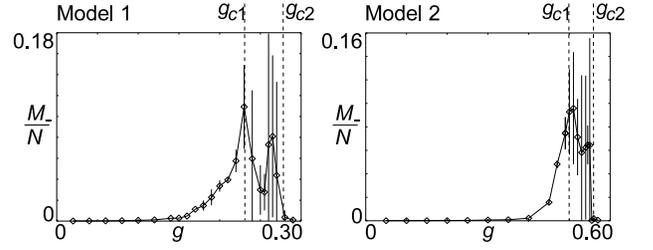}
\end{center}
\caption{The portion $M_{-}/N$
of the lattice occupied by the \emph{next} largest frequency cluster as a function of
coupling strength $g$. The maximum of $M_{-}/N$ corresponds to $g_{c1}$ and it drops almost
to zero at $g_{c2}$.
For Model 1 it is found that $g_{c1}\approx0.23$ and $g_{c2}\approx0.28$,
and for Model 2 $g_{c1}\approx0.50$ and $g_{c2}\approx0.56$.
Average and standard deviation of $7$ realizations of Model 1 and $3$ realizations
of Model 2 are shown for each $g$.}
\label{Rfig6}
\end{figure}

\begin{figure}
\begin{center}
\includegraphics[clip=true]{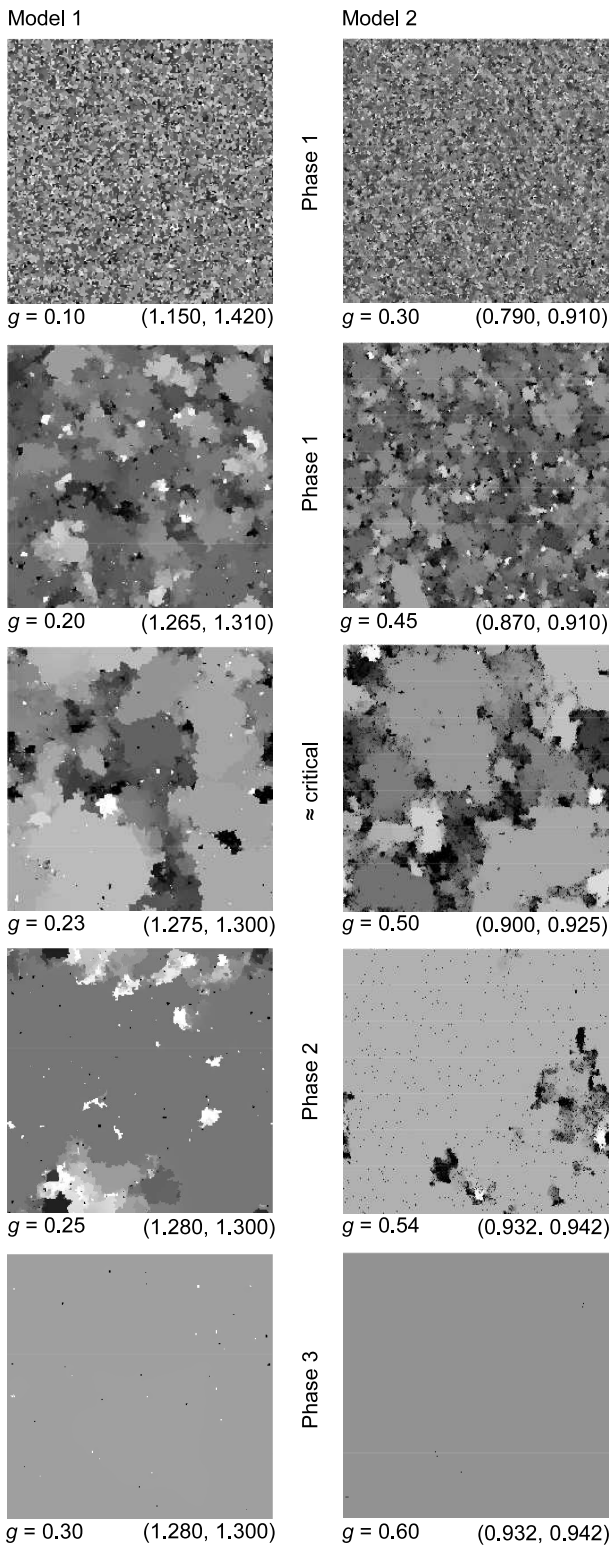}
\end{center}
\caption{Frequency landscapes $\Omega(\mathbf{r})$. Frequency is coded according to the bracket $(\omega_{\min}\;\omega_{\max})$.
An oscillator $k$ is colored black if $\Omega_{k}$ is less than $\omega_{\min}$ and white if
it is higher than $\omega_{\max}$.}
\label{Rfig7}
\end{figure}

I want to compare each theoretical prediction with numerical results from the two test models,
Model 1 [Eq. (\ref{model1})] and Model 2 [Eq. (\ref{model2})], in the case $D=2$.
To do so, it has to be demonstrated that there are critical points
in these models, and the critical coupling strengths $g_{c1}$ have to be identified.

Previously, two critical couplings $g_{c1}$ and $g_{c2}$ were found in Model 2 \cite{ptorus}.
The system seemingly becomes critical at $g=g_{c1}$ and almost perfect synchronization
settles at $g=g_{c2}$. There are still isolated oscillators that never fire for $g>g_{c2}$,
and thus they are not synchronized to the rest of the lattice \cite{strang}. The two critical couplings
separate three phases, \emph{phase 1} ($0\leq g < g_{c1}$), \emph{phase 2} ($g_{c1} < g < g_{c2}$),
and \emph{phase 3} ($g > g_{c2}$). These conclusions were reached by looking at the distribution
of cluster sizes. At $g=g_{c1}$, this distribution seems to obey a power law. This is true
in phase 2 also, if the macroscopic cluster is disregarded.
By estimating $g_{c1}$ and $g_{c2}$ for
different values of $N$, it was argued that the three phases are not a finite size effect,
but persist as $N\rightarrow\infty$. 

Simulations with Model 1 suggest that it behaves qualitatively in the same way. Instead of
showing cluster size distributions, we look in Fig. \ref{Rfig6} at the quantity $M_{-}/N$,
where $M_{-}$ is the size of the \emph{next} largest cluster. $M_{-}/N$ is expected to peak at $g_{c1}$.
Above $g_{c1}$ the largest, percolating cluster grows in size as $g$ increases further,
whereas the other clusters become smaller. At $g_{c2}$, $M_{-}/N$ should drop close to zero.
In this way, it is estimated that $g_{c1}\approx0.23$ and $g_{c2}\approx0.28$ for Model 1,
whereas $g_{c1}\approx0.50$ and $g_{c2}\approx0.56$ for Model 2 (Fig. \ref{Rfig6}).
Figure \ref{Rfig7} shows frequency landscapes $\Omega(\mathbf{r})$
in each of the three phases \cite{aboutmean}.

\begin{figure}
\begin{center}
\includegraphics[clip=true]{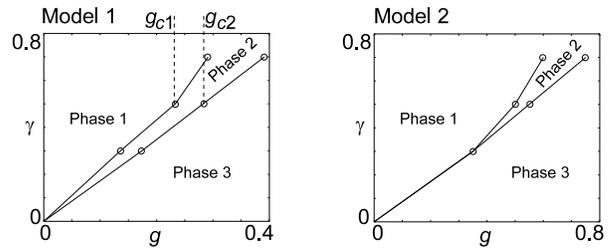}
\end{center}
\caption{Phase diagrams. The same quantity $M_{-}/N$
as in Fig. \ref{Rfig6} is used to identify $g_{c1}$ and $g_{c2}$. For model 1,
$\mathcal{D}_{\omega}$ is uniform with support $[1,1+\gamma]$. 
For model 2, $\mathcal{D}_{\omega^{-1}}$ is uniform with the same support.
Phase 1 is subcritical with microscopic frequency
clusters only. In phase 2, there is one macroscopic cluster. In phase 3, almost all
oscillators synchronize their freqencies. All three
phases seem to persist even if $\mathcal{D}_{\omega}$ has tails (see text).
For Model 2, it is impossible to resolve phase 2 in the data obtained with $\gamma=0.3$.
I cannot decide whether phase 2 extends down to the origin, or if there is a triple point.}
\label{Rfig8}
\end{figure}

For both models, phase 2 is rather narrow, but clearly distinguishable. To establish the
existence of phase 2 even more clearly, complementary
simulations were made for wider and narrower $\mathcal{D}_{\omega}$ (Fig. \ref{Rfig8}).

The hypothesis that will be tested is that the curve $g_{c1}(\sigma_{\omega}^{2})$ is identical
to the critical curve $S_{c}$, which is also the stable manifold of a critical fixed
point $\mathcal{E}^{*}$ (Fig. \ref{Rfig4}). Thus I identify $g_{c}=g_{c1}$.
It is a delicate question whether the entire phase 2 is critical.
In Ref. \cite{ptorus} I hypothesized that this is so, based on the cluster size distribution
and the temporal instability of clusters, even after very long times. The large
sample-to-sample fluctuations seen in Fig. \ref{Rfig6} further strenghten this idea.
The matter is discussed further below.

To test whether phase 3 exists even if $\mathcal{D}_{\omega}$ has tails,
Model 1 is simulated with Gaussian natural frequencies, with mean $\mu=0$ and
variance $\sigma_{\omega}^{2}=1/48$, i.e.
\begin{equation}
\mathcal{D}_{\omega}=\mathcal{N}(0,1/48).
\label{ndist}
\end{equation}
The variance is chosen to be equal to the
variance of the original, uniform $\mathcal{D}_{\omega}$. I estimate $g_{c1}=0.23$
and $g_{c2}=0.28$. Phase 3 is entered even if the oscillators with the most extreme
natural frequencies do not synchronize to the rest of the lattice.
Model 2 is simulated with the Rayleigh density function
\begin{equation}\left\{\begin{array}{ll}
\mathcal{D}_{\omega^{-1}}=4\pi(\omega^{-1}-1)e^{-2\pi(\omega^{-1}-1)^{2}}, & \omega^{-1}\geq 1 \\
\mathcal{D}_{\omega^{-1}}=0, & \omega^{-1}<1.
\end{array}\right.\end{equation}
In this model, it is found that $g_{c1}\approx 0.55$ and $g_{c2}\approx 0.75$. Phase 3 is entered
even if the slowest oscillators do not synchronize to the rest of the lattice.

I hypothesize that the transition to phase 3 is discontinuous, since the distribution
of cluster sizes seems to collapse discontiuously at $g_{c2}$. However, more detailed
studies are needed to establish the nature of this transition, and to be able to define
phase 3 precisely.

\subsection{Frequency correlations}

We have
$\textrm{E}[\Omega_{k}']=m_{\infty}'=m_{\infty}=\textrm{E}[\Omega_{k}]$ from Eqs.
(\ref{ensemble2condition}) and (\ref{samemean}) and
\begin{equation}
\textrm{Var}[\Omega_{k}']=\textrm{Var}[\Omega_{k}]+b^{-D}\sum_{k,k'\in j,\;k'\neq k}\textrm{Cov}[\Omega_{k},\Omega_{k'}]
\label{transvar}
\end{equation}
from Eq. (\ref{fstretch}) \cite{aboutmean}.
At a fixed point $\mathcal{E}^{*}$, the sum has to be zero for \emph{any} $b$ if
$\textrm{Var}^{*}[\Omega_{k}]<\infty$, which is the case treated here [Eq. (\ref{varexists})].
Therefore we must have
$\textrm{Cov}^{*}[\Omega_{k},\Omega_{k'}]=0$ for all $k\neq k$.
Introducing the pair correlation function
\begin{equation}
\Gamma_{\Omega}(r)=\textrm{Cov}[\Omega_{k},\Omega_{k'}]_{|k'-k|=r}/\textrm{Var}[\Omega_{k}],
\label{fcorrdef}
\end{equation}
it is concluded that
\begin{equation}
\Gamma_{\Omega}^{*}(r)\equiv 0,\ r\geq 1.
\label{nocorr}
\end{equation}
Note that even if $\Gamma_{\Omega}^{*}(r)\equiv 0$,
the $\Omega_{k}$s do not have to be \emph{independent} at a critical fixed point.
Rather, clusters of oscillators which run at the same frequency are expected (Fig. \ref{Rfig7}).

\begin{figure}
\begin{center}
\includegraphics[clip=true]{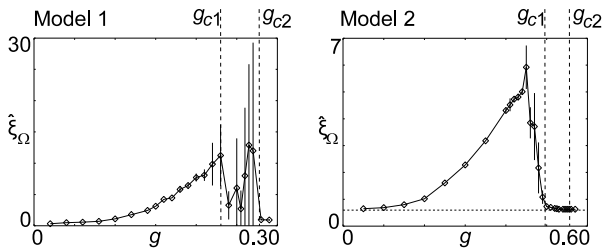}
\end{center}
\caption{The frequency correlation length $\hat{\xi}_{\Omega}$ [Eq. (\ref{fclength})].
According to Eq. (\ref{nocorr}), $\hat{\xi}_{\Omega}$ is expected to drop to zero
at $g=g_{c1}$. The same number of realizations as in Fig. \ref{Rfig6} are used.
Since linear interpolation is used, zero correlation between neighbors
gives $\hat{\xi}_{\Omega}=1-e^{-1}$ (dotted horizontal line).}
\label{Rfig9}
\end{figure}

Figure \ref{Rfig9} shows the correlation length $\hat{\xi}_{\Omega}$,
defined by the relation
\begin{equation}
\Gamma_{\Omega}(\hat{\xi}_{\Omega})=e^{-1}.
\label{fclength}
\end{equation}
(Note that this is not the standard way to define a correlation length, thus the hat symbol.
Normally it is defined as the rate of exponential fall-off at large $r$. C.f. Eq (\ref{corrform}).
The quantity $\hat{\xi}_{\Omega}$ is used here since it turned out to be more stable,
given the fluctuations of $\Gamma_{\Omega}(r)$.)
In Model 1, $\hat{\xi}_{\Omega}$ drops significantly just above the estimated value of
$g_{c1}$, with comparably small sample variations. In phase 2, $\hat{\xi}_{\Omega}$ seemingly
increases again, even if large variations make such a conclusion uncertain.
In Model 2, $\hat{\xi}_{\Omega}$ falls steeply towards zero just below $g_{c1}$,
and stay very close to zero for $g>g_{c1}$ with very small variations.

Thus, Model 2 supports the theory better than Model 1 does. However, the drop of $\hat{\xi}_{\Omega}$
in Model 1 may occur exactly at $g_{c1}$, given the uncertainty in its estimation
due to the large sample variations of the cluster sizes (Fig. \ref{Rfig6}).

No indications of negative correlations have been observed.

\subsection{Frequency distribution}
\label{freqdist}

\begin{figure}
\begin{center}
\includegraphics[clip=true]{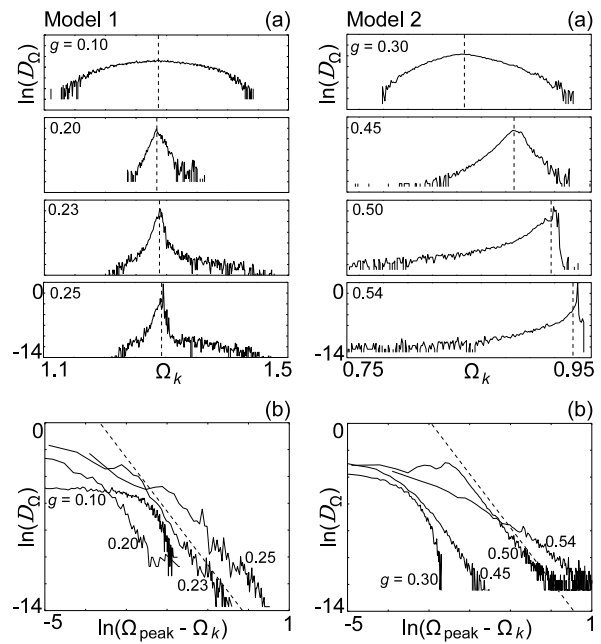}
\end{center}
\caption{Density functions $\mathcal{D}_{\Omega}$ of attained frequencies $\Omega_{k}$.
in Model 1 and Model 2. The vertical dashed lines in panels (a) show the mean frequency
$\mathrm{E}[\Omega_{k}]$. Panels (b) show the low frequency tails. The dashed lines correspond
to a relation $\mathcal{D}_{\Omega}(\Omega_{k})\propto(\Omega_{\mathrm{peak}}-\Omega_{k})^{-4}$.
The critical coupling strengths are $g_{c1}=0.23$ for Model 1 and $g_{c1}=0.50$ for Model 2.}
\label{Rfig10}
\end{figure}

Consider the density function $\mathcal{D}_{\Omega}$ of attained frequencies $\Omega_{k}$ \cite{aboutmean}.
I do not attempt to deduce $(\mathcal{D}_{\Omega})^{*}$ but discuss
some of its basic properties.

We have already concluded that assumption (\ref{finitevar})
implies that $\mathrm{E}^{*}[\Omega_{k}]$ and $\mathrm{Var}^{*}[\Omega_{k}]$ both exist.
Put differently, if the critical properties of Model 1 and Model 2 are to be the same
as those deduced for the fixed point $\mathcal{E}^{*}$, then $\mathrm{E}[\Omega_{k}]$ and
$\mathrm{Var}[\Omega_{k}]$ should exist at $g=g_{c1}$. Infinite $\mathrm{Var}[\Omega_{k}]$
at $g=g_{c1}$ and finite $\mathrm{Var}^{*}[\Omega_{k}]$ make necessary negative
frequency correlations [Eqs. (\ref{transvar}) and (\ref{fcorrdef})] along the critical
line $S_{c}$ (Fig.\ref{Rfig4}), which have not been observed in simulations.

At $g=g_{c1}$, the order parameter $r$ [Eq. (\ref{rdef})] becomes non-zero,
that is, a finite portion of the oscillators attain identical frequencies, so that
an infinitely high spike develops in $\mathcal{D}_{\Omega}$ as $g$ approaches $g_{c1}$
from below:
\begin{equation}
\lim_{g\rightarrow (g_{c1})_{\_}}\mathrm{Max}[\mathcal{D}_{\Omega}]=\infty.
\end{equation}
Let us use Eq. (\ref{fstretch}) to write
\begin{equation}
\tilde{\Omega}_{j}=\mathrm{E}[\Omega_{k}]+b^{-D/2}\sum_{k\in j}(\Omega_{k}-\mathrm{E}[\Omega_{k}]).
\label{numren}
\end{equation}
At a critical fixed point, the emerging spike should not move when $p_{b}$ is applied, so that
\begin{equation}
(\Omega_{\mathrm{peak}})^{*}=\mathrm{E}^{*}[\Omega_{k}],
\label{fixedpeak}
\end{equation}
where $\mathcal{D}_{\Omega}(\Omega_{\mathrm{peak}})=\mathrm{Max}[\mathcal{D}_{\Omega}]$.

Equation (\ref{numren}) can be used to renormalize a frequency landscape numerically, 
just like an Ising lattice can be renormalized by assigning the direction of
the block spin according to the majority rule. Ideally,
\begin{equation}
\lim_{b\rightarrow\infty}(\mathcal{D}_{\tilde{\Omega}})_{g_{c1}}=(\mathcal{D}_{\Omega})^{*},
\end{equation}
but there are numerical problems (apart from the dynamical instability).
First, small frequency gradients in a presumed cluster
are magnified by Eq. (\ref{numren}). Second, a block oscillator containing a cluster border
will not be part of the renormalized cluster, which distorts the renormalization of
cluster sizes for small and medium sized clusters, like those obtained in a simulation.
Basically, the problem is that the frequencies are continuous variables, whereas
spins are discrete.

\begin{figure}
\begin{center}
\includegraphics[clip=true]{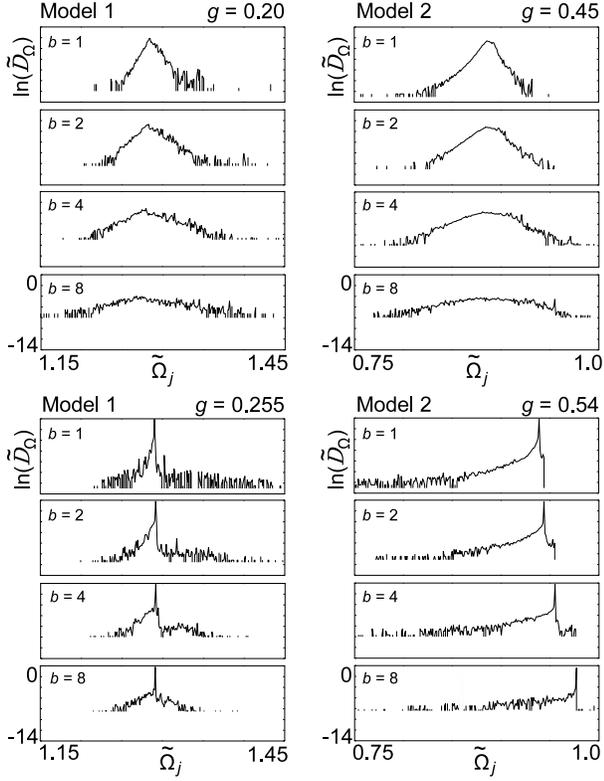}
\end{center}
\caption{Density functions $\tilde{\mathcal{D}}_{\omega}$ of attained
renormalized frequencies $\tilde{\Omega}_{j}$. Equation (\ref{numren})
is used for the numerical renormalization with different scale factors $b$.
In phase 1 ($g=0.20$ for Model 1 and $g=0.45$ for Models 2),
$\tilde{\mathcal{D}}_{\omega}$ approaches a normal distribution
correponding to a trivial fixed point at $g=0$. This does not seem to be
the case in phase 2 ($g=0.255$ for Model 1 and $g=0.54$ for Model 2).}
\label{Rfig11}
\end{figure}

Figure \ref{Rfig10} shows numerical estimations of $\mathcal{D}_{\Omega}$ for
Model 1 and Model 2. Panels (a) show that $\Omega_{\mathrm{peak}}\approx\mathrm{E}[\Omega_{k}]$,
at least for $g\leq g_{c1}$, indicating that Eq. (\ref{fixedpeak}) is fulfilled.
This is true even if $\mathcal{D}_{\Omega}$ is not symmetric about its mean.
However, in Model 2 for $g>g_{c1}$, it is clear that
$(\Omega_{\mathrm{peak}})>\mathrm{E}[\Omega_{k}]$.

Panels (b) in Fig. \ref{Rfig10} show the low frequency tails for each model
at different values of $g$. Just below and at $g=g_{c1}$ it seems that
\begin{equation}
\mathcal{D}_{\Omega}(\Omega_{k})\propto(\Omega_{\mathrm{peak}}-\Omega_{k})^{-\chi}
\label{ftail}
\end{equation}
with $\chi\approx 4$ (dashed lines), suggesting that the requirement that
$\mathrm{Var}[\Omega_{k}]$ exists at $g=g_{c1}$ is indeed fulfilled.
Regarding the high frequency tails, in Model 1 they seem to fall off quicker than
a power law for the largest frequencies for all $g$. In model 2, there seems to be
no tails close to and above $g_{c1}$. 

Figure \ref{Rfig11} shows numerical renormalizations in phases 1 and 2
using Eq. (\ref{numren}). The outcomes are qualitatively different,
indicating that different fixed points are approached in the two cases.
We have not been able to obtain convergence towards the presumed critical fixed point
density function
$\mathcal{D}_{\Omega}^{*}$ using a system close to $g=g_{c1}$.
Instead, the outcome is similar to that shown in phase 1,
although the normal distribution is approached more slowly (as $b$ increases).
The reason may be that, numerically, some positive correlations are still remaining
(Fig. \ref{Rfig9}).

The frequency correlation function $\Gamma_{\Omega}(r)$
drops close to zero in phase 2, but it seems that it never becomes negative. This means that
$\mathrm{Var}[\tilde{\Omega}_{j}]\geq \mathrm{Var}[\Omega_{k}]$.
Thus the flow under $R_{b}$ cannot go towards perfect synchronization
$(r=1)$, for which $\mathrm{Var}[\Omega_{k}]=0$, but it can reach states where the lattice
is synchronized except for isolated oscillators
with opposing frequencies. Such states are indeed seen at coupling strengths slightly
larger than $g_{c2}$ (Fig. \ref{Rfig7}). If there is a non-trivial fixed point corresponding
to such a state, then the outlier oscillators must have a fractal spatial distribution.
Otherwise they will "eat" the synhronized part of the lattice, and $r\rightarrow 0$
as $b\rightarrow\infty$. This effect is seen in Fig. \ref{Rfig11} as a (slightly)
decreasing height of the spike and an elevated baseline of outlier oscillators as $b$ increases.

\subsection{Cluster frequencies}

\begin{figure}
\begin{center}
\includegraphics[clip=true]{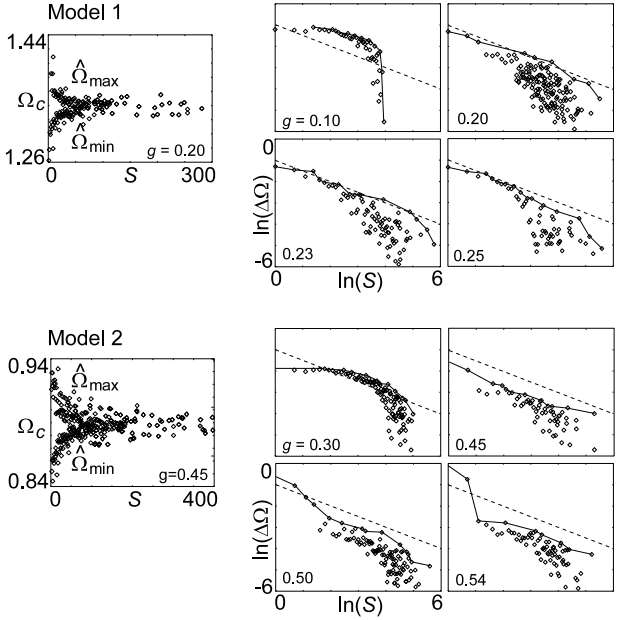}
\end{center}
\caption{$\hat{\Omega}_{\max}$ and $\hat{\Omega}_{\min}$ are the minimum and maximum frequencies
of clusters of size $S$ found numerically.
Let us write $\Delta\Omega=\hat{\Omega}_{\max}-\hat{\Omega}_{\min}$. Lower bounds on $\Delta\Omega_{\max}(S)$
are shown, assuming $d\Delta\Omega_{\max}/dS<0$, $\forall S$
(picewise linear, continuous curves). Dashed lines: predictions by Eq. (\ref{clusterfreq}) for a critical ensemble.
Data from 10 realizations of $\{\omega_{k}\}$ for each $g$.}
\label{Rfig12}
\end{figure}

Assume that the frequency
of a cluster $C$ with spatial size $S>>1$ is bounded by the inequality
\begin{equation}
|\Omega_{C}-m_{\infty}|<\Delta\Omega_{\max}(S),
\end{equation}
where $\Omega_{k}=\Omega_{C}$ whenever
$k\in C$ \cite{aboutmean}. For $t>>1$, choose $b<<S$ and apply $p_{b}$. We get $\tilde{S}=b^{-D}S$,
and $\Omega_{\tilde{C}}-m_{\infty}=b^{D/2}(\Omega_{C}-m_{\infty})$ from Eq. (\ref{fstretch}).
Consequently, $\Delta\Omega_{\max}'(b^{-D}S)=b^{D/2}\Delta\Omega_{\max}(S)$, and at a
fixed point we get
\begin{equation}
\Delta\Omega_{\max}^{*}(S)\propto S^{-1/2}.
\label{clusterfreq}
\end{equation}
Thus, the cluster frequencies vary less and less as their sizes increase.

Figure \ref{Rfig12} shows comparisons between the theoretical prediction in Eq. (\ref{clusterfreq})
and numerical data. The numerical difference $\hat{\Omega}_{\max}-\hat{\Omega}_{\min}$
as a function of $S$ is studied rather than the differences $\hat{\Omega}_{\max}-\hat{m}_{\infty}$ or
$\hat{m}_{\infty}-\hat{\Omega}_{\min}$, where $\hat{m}_{\infty}$ is a numerical estimation of
$m_{\infty}$. The reason is that I want to estimate as few quantities as possible.
Especially for large $S$, the latter differences are very sensitive to the choice of $\hat{m}_{\infty}$.
Note however, that if these differences are used, data is obtained that support Eq. (\ref{clusterfreq})
to the same extent as the data shown in Fig. \ref{Rfig12} does. Around $g_{c1}$ and in phase 2,
the agreement with theory is reasonable, given the fluctuations in the data. This gives
further support to the idea that the entire phase 2 is critical.
For Model 2, the large values of $\Delta\Omega$ for $g=0.50$ and $0.54$
for $S=1$ and $S=2$ are due to the existence
of isolated oscillators which are transiently suppressed, with $\Omega_{k}\approx 0$ \cite{ptorus,strang}.

\subsection{Frequency transient}

\begin{figure}
\begin{center}
\includegraphics[clip=true]{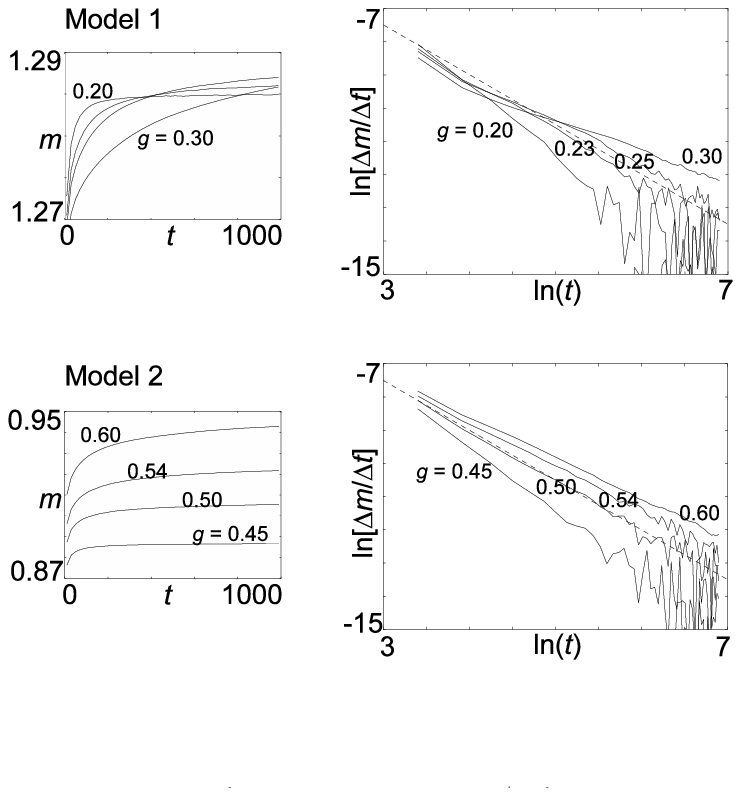}
\end{center}
\caption{Transient of mean frequency $m(t)$. Lattice size $2000\times 2000$.
The dashed lines correspond to the prediction by Eq. (\ref{fixtrans}) for a critical ensemble.
Good data quality enables use of $dm/dt$, so that no asymptotic values
have to be estimated (C.f. Fig. \ref{Rfig14}).}
\label{Rfig13}
\end{figure}

From Eq. (\ref{btrans}) we get
$\tilde{m}(t')=\mathrm{E}[d\tilde{\phi}_{j}/dt']=
b^{D/2}m(t)-(b^{D/2}-1)m_{\infty}$.
At a fixed point we have $m^{*}(b^{-D/2-1}t)-m_{\infty}^{*}=b^{D/2}(m^{*}(t)-m_{\infty}^{*})$
with solution
\begin{equation}
m^{*}(t)-m_{\infty}^{*}\propto t^{-D/(D+2)}.
\label{fixtrans}
\end{equation}

Close to the critical couplings $g_{c1}\approx 0.23$ (Model 1) and $g_{c1}\approx 0.50$ (Model 2),
the agreement with Eq. (\ref{fixtrans}) is excellent in the data shown in Fig. \ref{Rfig13}.
In the double-logarithmic plots, there is a tendency to a gradual increase of the slope
as $g$ increases, suggesting that the scaling expressed in Eq. (\ref{fixtrans}) only
applies at $g_{c1}$, and not in the entire phase 2. This in turn suggests that phase 2
is \emph{not} critical, at least that it is not attracted to the critical fixed point $\mathcal{E}^{*}$
described in this paper.

\subsection{Mean frequency for finite $N$}

Taking the ensemble mean of Eq. (\ref{transomega}) gives
$\mathrm{E}[\tilde{\omega}_{j}]-m_{\infty}=
b^{D/2}(\mathrm{E}[\underline{m}_{j}]-m_{\infty})$.
We may write $m_{\infty}=m_{\infty}(N)$, and then have
$\mathrm{E}[\underline{m}_{j}]=m_{\infty}(b^{D})$.
If we first let the size of the whole lattice go to infinity and then set $b^{D}=N$, we get
\begin{equation}
m_{\infty}(N)-m_{\infty}(\infty)=N^{-1/2}\left\{\mathrm{E}[\tilde{\omega}_{j}]-m_{\infty}(\infty)\right\}.
\end{equation}
Taking the ensemble mean of Eq. (\ref{tildetrans}) in the limits $N\rightarrow\infty$
and $t\rightarrow\infty$, we get $\tilde{m}_{\infty}(\infty)=\mathrm{E}[\tilde{\omega}_{j}]+\mathrm{E}[\tilde{\underline{h}}_{j}]$.
Using Eq. (\ref{samemean}), we may therefore write
\begin{equation}
m_{\infty}(\infty)-m_{\infty}(N)=N^{-1/2}\mathrm{E}[\tilde{\underline{h}}_{j}].
\label{sizefreq1}
\end{equation}
At a critical fixed point $\mathcal{E}^{*}$,
$\mathrm{E}^{*}[\tilde{\underline{h}}_{j}]=\mathrm{E}^{*}[\underline{h}_{k}]$ for all $b$ (or $N$)
and therefore $m_{\infty}(N)-m_{\infty}(\infty)\propto N^{-1/2}$ for all $N$. At a critical ensemble
attracted to $\mathcal{E}^{*}$, we have
$\mathrm{E}^{*}[\tilde{\underline{h}}_{j}]\rightarrow\mathrm{E}^{*}[\underline{h}_{k}]$ as $b\rightarrow\infty$,
so that
\begin{equation}
m_{\infty}(N)-m_{\infty}(\infty)\propto N^{-1/2},\ N>>1.
\label{sizefreq}
\end{equation}
For odd coupling [Eq. (\ref{odd})], we have
$\mathrm{E}[\tilde{\underline{h}}_{j}]=\mathrm{E}[\underline{h}_{k}]=0$ and
$m_{\infty}(N)=\mathrm{E}[\omega_{k}]$ for all $N$.
This corresponds to zero constant of proportionality in Eq. (\ref{sizefreq}).

\begin{figure}
\begin{center}
\includegraphics[clip=true]{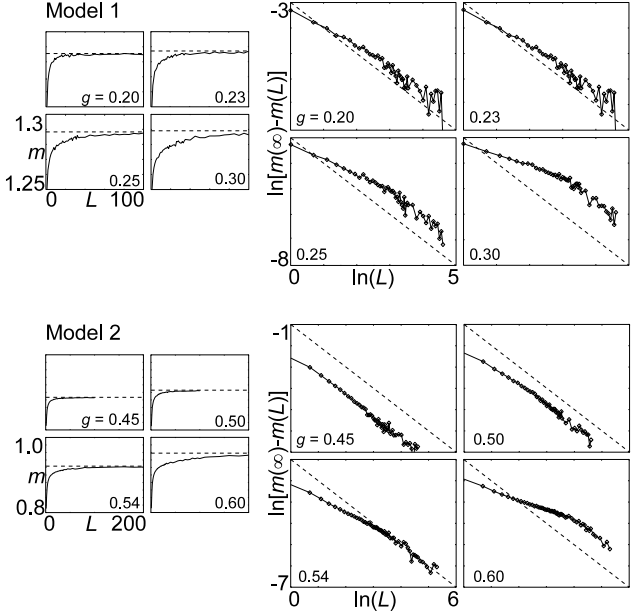}
\end{center}
\caption{Mean frequency $m(t)$ at time $t=1000$ versus $L=\sqrt{N}$.
Averages of up to $400 000/N$ realizations.
Estimated values of $m(\infty)$ (dashed horizontal lines) provide good agreement
with Eq. (\ref{sizefreq}) for all $g$ (dashed diagonal lines).}
\label{Rfig14}
\end{figure}

For sub-critical ensembles it is expected that
$\lim_{b\rightarrow\infty}\mathrm{E}[\tilde{\underline{h}}_{j}]=0$,
and for super-critical ensembles that
$\lim_{b\rightarrow\infty}\mathrm{E}[\tilde{\underline{h}}_{j}]=\infty$.
In both cases, Eq. (\ref{sizefreq}) cannot be expected to hold as $N\rightarrow\infty$.

Nevertheless, the data in Fig. \ref{Rfig14} is consistent with Eq. (\ref{sizefreq})
for all shown $g$. However, crossover to another scaling for larger $L=\sqrt{N}$ cannot be excluded.
For both models, the asymptotic behavior is reached for larger $L$ for higher values of $g$.
Due to poor data quality (C.f. Fig. \ref{Rfig13}),
estimated values of $m(\infty)$ have to be relied upon, chosen to get curves
in the double-logarithmic plots that are as straight as possible for large $L$.
Another possible source of error is that I had to compute the mean frequencies
at a rather small time $t=1000$ due to limited computational resources. However,
tests with smaller and larger $t$ indicate that this is not crucial.

\subsection{Correlations of interactions}

\begin{figure}
\begin{center}
\includegraphics[clip=true]{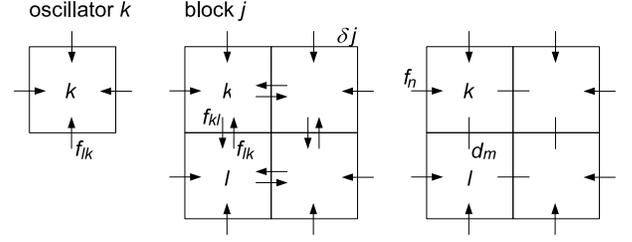}
\end{center}
\caption{The interaction $h_{k}$ is a sum of the $2D$ coupling functions $f_{lk}$.
In the same way, the interaction $\tilde{h}_{j}$ of block $j$ can be decomposed into a sum of
$2Db^{D-1}$ border terms $f_{n}$ and $Db^{D}(1-b^{-1})$ interior tems $d_{m}=f_{lk}+f_{kl}$.}
\label{Rfig15}
\end{figure}

Let us turn to the renormalization of the interactions $h_{k}$. 
It turns out to be useful to decompose $h_{k}$ into the coupling functions $f_{lk}$,
and to define the asymmetry function
\begin{equation}
d_{m}(x,y)\equiv f_{lk}(y,x)+f_{kl}(x,y),
\label{defd}
\end{equation}
where $m$ is the edge connecting oscillators $l\in j$ and $k\in j$.
Let $n$ be a directed edge across $\delta j$ (Fig. \ref{Rfig15})
and let us write $f_{n}=f_{lk}$, where $l\notin j$ and $k\in j$. We then have
\begin{equation}
\underline{\tilde{h}}_{j}=b^{-D/2}g[\sum_{n}\underline{f}_{n}+
\sum_{m}(\underline{d}_{m}-\underline{d}_{m}^{\prec})],
\end{equation}
where $d_{m}^{\prec}(x,y)=f_{lk}^{\prec}(x,y)+f_{kl}^{\prec}(y,x)$.
To make the following expressions more compact, let
\begin{equation}
\Delta \underline{d}_{m}=\underline{d}_{m}-\underline{d}_{m}^{\prec}
\end{equation}
be the mean increase of $d_{m}$ as block $j$ is connected to its neighbor block oscillators.
For odd coupling, i.e.
\begin{equation}
f_{lk}(y,x)\equiv-f_{kl}(x,y), \ \forall lk,
\label{odd}
\end{equation}
we have $d_{m}\equiv d_{m}^{\prec}\equiv 0$. An example is the Kuramoto model
$f_{lk}(x,y)=\sin[2\pi(x-y)]$.

Information about critical behavior can be gained by comparing
moments of the original and renormalized interactions: $\mathrm{E}[\underline{h}_{k}]$, $\mathrm{E}[\underline{h}_{k}^{2}]$, and so on.
A comparison between $\mathrm{E}[\underline{h}_{k}]$ and
$\mathrm{E}[\tilde{\underline{h}}_{j}]$ just leads us back to Eq. (\ref{sizefreq1}).
Below, I focus instead on $\mathrm{Var}[\underline{h}_{k}]$
and $\mathrm{Var}[\underline{\tilde{h}}_{j}]$,
from which information can be gained of two-point correlations of the interaction.
At this point we make use of assumption (\ref{finitevar}). We may then write
\begin{equation}
\mathrm{Var}[\underline{\tilde{h}}_{j}]= b^{-D}\sum_{k,k'\in j}\mathrm{Cov}[\underline{h}_{k}-\underline{h}_{k}^{\prec},\underline{h}_{k'}-\underline{h}_{k'}^{\prec}],
\end{equation}
or, upon decomposition,
\begin{equation}\begin{array}{lll}
\mathrm{Var}[\underline{\tilde{h}}_{j}] & = &
b^{-D}g^{2}\sum_{n,n'}\mathrm{Cov}[\underline{f}_{n},\underline{f}_{n'}]+\\
&& b^{-D}g^{2}\sum_{n,m}\mathrm{Cov}[\underline{f}_{n},\Delta \underline{d}_{m}]+\\
&& b^{-D}g^{2}\sum_{m,m'}\mathrm{Cov}[\Delta \underline{d}_{m},\Delta \underline{d}_{m'}]\\
& = & S_{1}+S_{2}+S_{3}.
\end{array}\label{terms}\end{equation}
In the following, three correlation functions will be used:
\begin{equation}\begin{array}{lll}
\Gamma_{\underline{f}}(r) & = & \lim_{t\rightarrow\infty}\frac{\textrm{Cov}
[\underline{f}_{lk},\underline{f}_{l'k'}]_{|l'k'-lk|=r}}
{\textrm{Var}[\underline{f}_{lk}]}\\
\Gamma_{\underline{f}\Delta \underline{d}}(r) & = & \lim_{t\rightarrow\infty}\frac{\textrm{Cov}
[\underline{f}_{n},\Delta\underline{d}_{m}]_{|n-m|=r}}
{\textrm{Cov}[\underline{f}_{n},\Delta\underline{d}_{m}]_{|n-m|=1}}\\
\Gamma_{\Delta\underline{d}}(r) & = & \lim_{t\rightarrow\infty}\frac{\textrm{Cov}
[\Delta\underline{d}_{m},\Delta\underline{d}_{m}]_{|m-m'|=r}}
{\textrm{Var}[\Delta\underline{d}_{m}]}.
\end{array}\end{equation}
Note that $f_{lk}$ has a direction of influence $l\rightarrow k$, and that $d_{m}$ is vertical or horizontal.
Correlations between different types of pairs should therefore be separated, and summed up to yield the
covariances in Eq. (\ref{terms}).
Numerically, only parallel pairs are considered.

A necessary fixed point condition is
$\mathrm{Var}^{*}[\underline{\tilde{h}}_{j}]=\mathrm{Var}^{*}[\underline{h}_{k}]$
for any $b$ as $t\rightarrow\infty$,
or, in particular, at a non-trivial fixed point,
\begin{equation}
\lim_{b\rightarrow\infty}\lim_{t\rightarrow\infty}\mathrm{Var}^{*}[\underline{\tilde{h}}_{j}]=const.>0.
\end{equation}

\begin{figure}
\begin{center}
\includegraphics[clip=true]{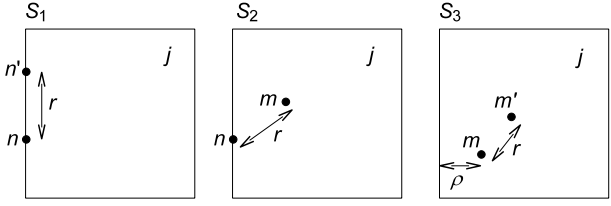}
\end{center}
\caption{Pairs of interactions and distances in a block-oscillator $j$, used in the expressions for
$S_{1}$, $S_{2}$, and $S_{3}$ in Eqs. (\ref{terms}), (\ref{s1}), (\ref{s2}), and (\ref{s3}).}
\label{Rfig16}
\end{figure}

We may write
\begin{equation}
S_{1}\propto b^{-D}\int_{1}^{b}N_{1}(r)\Gamma_{\underline{f}}(r)dr,
\label{s1}
\end{equation}
where
\begin{equation}N_{1}(r)=\mathcal{O}(b^{D-1}r^{D-2})
\label{n1}
\end{equation}
is the number of pairs $nn'$ at distance $r$ (Fig. \ref{Rfig16}).
It follows that
\begin{equation}
\lim_{b\rightarrow\infty}S_{1}=const.>0\Leftrightarrow
\Gamma_{\underline{f}}^{*}(r)\propto r^{2-D}.
\end{equation}

Similarly,
\begin{equation}
S_{2}\propto b^{-D}\int_{1}^{b}N_{2}(r)\Gamma_{\underline{f}\Delta \underline{d}}(r)dr,
\label{s2}
\end{equation}
where
\begin{equation}
N_{2}(r)=\mathcal{O}(b^{D-1}r^{D-1})
\end{equation}
is the number of pairs $nm$ at distance $r$ (Fig. \ref{Rfig16}). Therefore it is expected that
\begin{equation}
\lim_{b\rightarrow\infty}S_{2}=const.>0 \Leftrightarrow
\Gamma_{\underline{f}\Delta \underline{d}}^{*}(r)\propto r^{1-D}.
\end{equation}

Turning to $S_{3}$, we may write
\begin{equation}
\mathrm{Cov}[\Delta\underline{d}_{m},\Delta\underline{d}_{m'}]=\psi(\rho)\Gamma_{\Delta \underline{d}}(r),
\label{s3}
\end{equation}
where $\rho$ is the smallest distance from any of the two edges $m$ or $m'$ to $\delta j$,
and $r$ is the distance between $m$ and $m'$ (Fig. \ref{Rfig16}). The function
\begin{equation}
\psi(\rho)=\mathrm{Var}[\Delta\underline{d}_{m}]
\label{psidef}
\end{equation}
measures how the lattice that surrounds block $j$, in the mean,
changes $\underline{d}_{m}$ at distance $\rho$ from $\delta j$,
so that we have
\begin{equation}\begin{array}{c}
\psi(1)>0\\
\lim_{\rho\rightarrow\infty}\psi(\rho)=0.
\end{array}\end{equation}
We may then write
\begin{equation}
S_{3}\propto b^{-D}\int_{1}^{b/2}\psi(\rho)\int_{1}^{b-2\rho}N_{3}(\rho,r)\Gamma_{\Delta \underline{d}}(r)dr d\rho.
\end{equation}
Here,
\begin{equation}
N_{3}(\rho,r)=\mathcal{O}(\rho^{D-1}r^{D-1})
\end{equation}
is the number of pairs $mm'$ for given $\rho$.
In a critical fixed point ensemble, it is expected that
\begin{equation}
\psi^{*}(\rho)\propto \rho^{-\alpha}
\label{criticalpsi}
\end{equation}
and therefore we get
\begin{equation}
\lim_{b\rightarrow\infty}S_{3}=const.>0\Leftrightarrow\Gamma_{\Delta \underline{d}}^{*}(r)\propto r^{\alpha-D},
\label{s3cond}
\end{equation}
provided $\alpha\neq 1$.

\begin{figure}
\begin{center}
\includegraphics[clip=true]{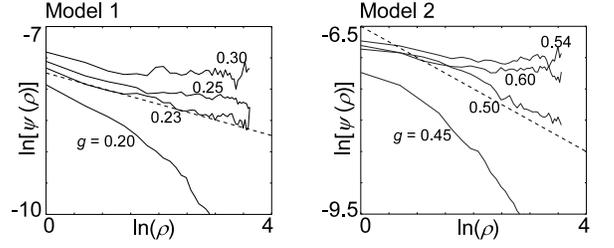}
\end{center}
\caption{$\psi(\rho)=\mathrm{Var}[\Delta\underline{d}_{m}]$ as a function of the distance $\rho$
from $\delta j$ to the edge $m$ (Fig. \ref{Rfig16}). For both models, it seems that
$\psi(\rho)\propto \rho^{-\alpha}$ at $g_{c1}$. For Model 1, $\alpha\approx 1/4$,
and for Model 2, $\alpha\approx 1/2$. Lattice size: $140\times 140$.
Block size: $70\times 70$. Transient time: $t=5000$. For each of $10$ realizations of
$\{\omega_{k}\}$, $20$ initial conditions $\phi(0)$ were used.}
\label{Rfig17}
\end{figure}

I have not been able to deduce the value of $\alpha$ from first principles.
Figure \ref{Rfig17} shows numerical estimations of $\psi(r)$ for Model 1 and Model 2. It seems that
$\alpha=1/4$ for Model 1, and $\alpha=1/2$ for Model 2, and thus that it is a non-universal,
model dependent critical exponent. The scaling form (\ref{criticalpsi}) seems to apply only
at $g=g_{c1}$, suggesting that phase 2 is not critical.

It can be argued that since $\underline{d}_{m}$ and $\Delta \underline{d}_{m}$
are linear combinations of $\underline{f}_{lk}$ and $\underline{f}_{lk}^{\prec}$,
$\Gamma_{\underline{f}}$, $\Gamma_{\underline{d}}$, $\Gamma_{\underline{f}\Delta\underline{d}}$ and $\Gamma_{\Delta\underline{d}}$,
should have the same functional form for $r>>1$. Then,
\begin{equation}
\Gamma_{\underline{f}}^{*}\propto\Gamma_{\underline{d}}^{*}\propto r^{-\beta}
\end{equation}
with
\begin{equation}\begin{array}{lcllll}
\beta=D-2, &  odd & f_{lk} & & & (case\;1)\\
\beta=D-1, & other & f_{lk}, & \alpha\geq 1 & & (case\;2) \\
\beta=D-\alpha, & other & f_{lk}, & \alpha<1 & & (case\;3).
\end{array}
\label{criticaltable}
\end{equation}
The terms $S_{1}$, $S_{2}$ and $S_{3}$ are responsible for criticality in the three cases, respectively.
In case 1,
\begin{equation}\begin{array}{lll}
S_{1}^{*} & = & const.>0\\
S_{2,3}^{*} & = & 0.
\end{array}\end{equation}
[In fact, $S_{2,3}=0$ for all ensembles since $d_{m}(x)\equiv 0$.] In case 2,
\begin{equation}\begin{array}{lll}
\lim_{b\rightarrow\infty}S_{2}^{*} & = & const.>0\\
\lim_{b\rightarrow\infty}S_{1,3}^{*} & = & 0,
\end{array}\end{equation}
and in case 3,
\begin{equation}\begin{array}{lll}
\lim_{b\rightarrow\infty}S_{3}^{*} & = & const.>0\\
\lim_{b\rightarrow\infty}S_{1,2}^{*} & = & 0.
\end{array}\end{equation}
In cases 1 and 2, critical behavior is ruled out below $D=3$ and
$D=2$, respectively, since correlations must decay with $r$.
In case 1, the result $D_{c}\geq 2$ by Daido \cite{daido} is regained.

\begin{figure}
\begin{center}
\includegraphics[clip=true]{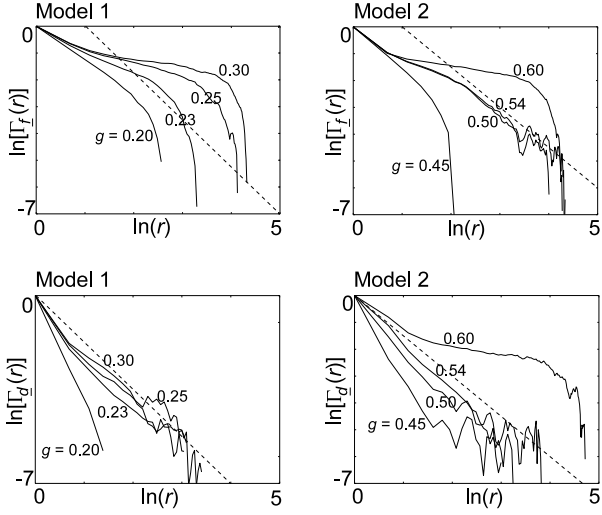}
\end{center}
\caption{Pair-correlations of $\underline{f}$
(left) and $\underline{d}$ (right). According to Eq. (\ref{criticaltable}) and the estimations
of $\alpha$ in Fig. \ref{Rfig17}, it is expected that $\beta=-7/4$ for Model 1, and $\beta=-3/2$ for Model 2
in critical ensembles (dashed lines). Ten $\phi(0)$ were used for each $g$ to estimate the
initial condition mean. In Model 2, a time average of each $f_{lk}$ was calculated
during $\Delta t=100$, due to its pulse-like nature.
The choice of $\Delta t$ did not affect the results.}
\label{Rfig18}
\end{figure}

The numerical results in Fig. \ref{Rfig17} suggest that $\alpha<1$ for both Model 1
and Model 2, and thus that the term $S_{3}$ is responsible for criticality in both
models. However, since the estimated values of $\alpha$ differ, it is possible
that there are other non-odd models whith $\alpha\geq 1$, in which case $S_{2}$
becomes the crucial term.

Figure \ref{Rfig18} shows numerical estimations of $\Gamma_{\underline{f}}(r)$
and $\Gamma_{\underline{d}}(r)$.
Looking at $\Gamma_{\underline{d}}(r)$,
the data is consistent with the combined theoretical and numerical predictions
[Eq. (\ref{criticaltable}) and Fig. \ref{Rfig17}] in both Model 1 and Model 2.
Looking at $\Gamma_{\underline{f}}(r)$, the data are consistent with theory
only in Model 2.

The reason for this discrepancy between numerical data and theory in Model 1 is likely to be found
in the fact that $\Gamma_{\underline{f}}(r)$ is less well-behaved than $\Gamma_{\underline{d}}(r)$
for finite lattice sizes. The phase fields $\phi(x,y)$ become more and more well-ordered
as $g$ increases, containing just a few foci or spirals as phase 3 is approached
(at the present lattice size) \cite{strang}. Therefore the phase waves tend to move in opposite
directions at opposite ends of the lattice, giving rise to negative correlations of $\underline{f}$
at large distances.
This dependence on the wave direction
of the correlations is eliminated by the definition of $d$ [Eq. (\ref{defd})].

\begin{figure}
\begin{center}
\includegraphics[clip=true]{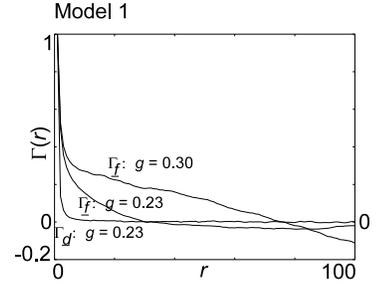}
\end{center}
\caption{Illustration of the fact that numerical estimations of $\Gamma_{\underline{f}}(r)$ 
are less well-behaved than those of $\Gamma_{\underline{d}}(r)$. For a given finite lattice size,
$\Gamma_{\underline{f}}(r)$ becomes more ill-behaved when $g$ increases (see text).
In Model 2, a time average of each $f_{lk}$ was calculated
during $\Delta t=100$. (C.f. Fig. \ref{Rfig18}).}
\label{Rfig19}
\end{figure}

This problem is illustrated for Model 1 in Fig. \ref{Rfig19}. Close to criticality, for $g=0.23$,
$\Gamma_{\underline{d}}(r)$ converges nicely towards zero as $r$ increases, whereas
$\Gamma_{\underline{f}}(r)$ drops significantly below zero, and then fluctuate, at least
up to $r=150$. (This is the maximum $r$ considered, since the lattice size is $300\times 300$.)
This effect is more prominent for larger $g$ as seen in the estimation of
$\Gamma_{\underline{f}}(r)$ for $g=0.30$.
The zero-crossings of $\Gamma_{\underline{f}}(r)$ is the reason why the curves drop
sharply in the double-logarithmic plots in Fig. \ref{Rfig18}.

\subsection{The correlation length}

Let us analyze $\Gamma_{\underline{f}}$ close to a critical fixed point
in a subcritical ensemble, and make the standard ansatz
\begin{equation}
\Gamma_{\underline{f}}(r)=cr^{-\beta}e^{-r/\xi(\Delta g)},
\label{corrform}
\end{equation}
where
\begin{equation}
\Delta g=(g-g^{*})/g^{*}.
\end{equation}
As discussed below, subcriticality is expected only for $g<g^{*}$.
It is therefore assumed that $\Delta g\leq 0$. We may write $\mathrm{Var}[\underline{f}_{lk}]=F(g)$.
Assuming that $dF/dg\neq 0$ at $g=g^{*}$, we have
\begin{equation}
\Delta g\propto\mathrm{Var}^{*}[\underline{f}_{lk}]-\mathrm{Var}[\underline{f}_{lk}]
\end{equation}
for small enough $\Delta g$. Consider the case of odd coupling.
From the expression (\ref{s1}) for $S_{1}$ we get
\begin{equation}
\Delta g'\propto b^{-1}\int_{1}^{b}r^{D-2}[\Gamma_{\underline{f}}^{*}(r)-\Gamma_{\underline{f}}(r)]dr.
\end{equation}
Taylor expanding the exponential part of $\Gamma_{\underline{f}}$ gives $\Delta g'\propto b\xi^{-1}$.
Using $\xi'=\xi/b$, specifying $\xi=\xi_{\underline{f}}$,
it is seen that the correlation length of the initial condition mean of the coupling $f_{lk}$
diverges according to
\begin{equation}
\xi_{\underline{f}}\propto \Delta g^{-1},
\label{correl}
\end{equation}
for small enough $|\Delta g|$ if $\Delta g<0$. A similar calculation
gives the same result in the cases of non-odd coupling, using the expressions for $S_{2}$ and $S_{3}$.

Indeed, simulations suggest that $\xi_{\underline{f}}$ diverges at $g=g_{c1}$, but
unfortunately I have not been able to obtain good enough data to test relation (\ref{correl}).
The fluctuations in the estimated $\xi_{\underline{f}}$ are too large close to $g_{c1}$.
(I used up to three realizations of $\{\omega_{k}\}$ for each $g$, and for each
$\{\omega_{k}\}$, ten $\phi(0)$ were used to estimate the initial condition mean).
It was not possible to use data from estimations of $\Gamma_{\underline{d}}$ either,
since it drops close to zero for too small $r$ to be able to resolve its functional form.

\subsection{Direction of the renormalization flow}
\label{rflow}

In Fig. \ref{Rfig17} the exponent $\alpha$ is estimated in the relation $\psi(\rho)\propto \rho^{-\alpha}$,
[Eqs. (\ref{psidef}) and (\ref{criticalpsi})], that is expected to hold in a critical ensemble.
Let us call these estimations $\alpha_{1}$ and $\alpha_{2}$
for models 1 and 2, respectively. In phases 2 and 3, $\psi(\rho)$ clearly falls off slower than this.
Figure \ref{Rfig18} shows that in phases 2 and 3, $\Gamma_{\underline{f}}$ and $\Gamma_{\underline{f}}$
falls off as $r^{-(D-\alpha_{1})}$ (Model 1),  $r^{-(D-\alpha_{2})}$ (Model 2), or possibly slower.
Taken together, these observations suggest that condition (\ref{s3cond}) is violated in phases 2 and 3,
and that $\lim_{b\rightarrow\infty}S_{3}=\infty$. This in turn means that
$\lim_{b\rightarrow\infty}\mathrm{Var}[\underline{\tilde{h}}_{j}]=
\lim_{b\rightarrow\infty}\mathrm{Var}[\underline{h}_{k}']=\infty$,
and that the renormalization flow goes in the direction of increasing $g$ for $g>g_{c1}$
(Fig. \ref{Rfig20}). That the flow goes towards $g=0$ for $g<g_{c}$ becomes clear from a similar argument.
I have mentioned the possibility that the entire phase 2 is critical, and that it is attracted to the
critical fixed point $\mathcal{E}^{*}$. Some numerical results favor such an interpretation
(see Figs. \ref{Rfig9}, \ref{Rfig12}, \ref{Rfig13}, \ref{Rfig14}, \ref{Rfig18}, and also
Ref. \cite{ptorus}). However, based on the combined numerical and theoretical argument given above,
I hypothesize that this is not so.

Referring to the discussion in section \ref{freqdist}, it seems that the renormalization
flow in phase 2 cannot approach states with $r=1$. Therefore it is probable that phase 2
is invariant under $R_{b}$. There may be a second, attractive
fixed point with $\mathrm{Var}[\Omega_{k}]=\infty$ somewhere along the line separating phases 2 and 3,
possibly at infinity where $g\rightarrow\infty$ or $\sigma_{\omega}^{2}\rightarrow\infty$.

Correlation functions seem to decay as a power law or slower
for all $g>g_{c1}$. In fact, Eqs. (\ref{s1}), (\ref{s2}) and (\ref{s3}) predict that finite correlations lengths
are excluded for $g>g_{c1}$, since whenever $\Gamma_{\underline{f}}$ has an exponential factor,
$\lim_{b\rightarrow\infty}S_{1-3}=0$.
This corresponds to a renormalization flow towards $g=0$ ($\mathrm{Var}[\underline{\tilde{h}}_{j}]=0$),
which can be expected only for $g<g_{c1}$. Therefore, phases 2 and 3 must be considered
\emph{supercritical}.

\section{Discussion}

\begin{figure}
\begin{center}
\includegraphics[clip=true]{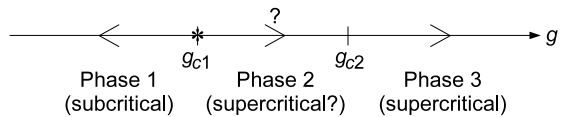}
\end{center}
\caption{Hypothetical flow under $R_{b}$ in a one-dimensional projection of $\{\mathcal{E}\}$.
The critical coupling strength $g_{c1}$ is supposed to belong to the stable manifold
$S_{c}$ of the critical fixed point $\mathcal{E}^{*}$ (c.f. Fig. \ref{Rfig2}).
It is speculated that the flow does not pass $g_{c2}$, i.e. that phase 2 is invariant.
See text for explanation.}
\label{Rfig20}
\end{figure}

In this paper, I present a real-space renormalization transformation for oscillator lattices
with quenched disorder. The transformation acts on ensembles of lattices
and predicts the behavior of ensemble averaged quantities.
It is assumed that the variance of the natural and attained frequencies exists,
but it should be possible to generalize the theory.
A bold hypothesis is that \emph{if} a system of form (\ref{model}) is critical for some
parameter values, then the critical behavior is given by the critical fixed point $\mathcal{E}^{*}$
desribed in this paper. At its present stage, the theory cannot be used to decide
\emph{whether} a given system posesses a critical phase transition. However, lower
bounds on critical dimensions for different classes of systems are given.

In this respect, the crucial difference between odd and non-odd coupling stands out clearly in the analysis.
For non-odd coupling, macroscopic synchronization cannot be ruled out for any dimension $D\geq 1$,
whereas for odd coupling it is necessary that $D\geq 3$.
Perfectly odd coupling must be regarded as a non-generic special case,
except for particular problems that can be mapped onto Kuramoto-like models,
such as Josephson junction arrays \cite{josephson}.

The merits of the approach are that it is simple, that it applies to a broad class
of systems, that several predictions
about critical behavior can be extracted, and that it is potentially exact.
Most of the predictions have been tested numerically with two structurally different
two-dimensional models. The agreement with theory ranges from acceptable
to very good. The drawback of the approach is that the theory must be considered heuristic
at its present stage. Its full potential and its mathematical foundation should be clarified.

My experience is that it is computationally demanding to get good numerical data
to compare with theory.
Large oscillator lattices [$\mathcal{O}(10^{5})$ oscillators] and long simulation times
[$\mathcal{O}(10^{5})$ periods of oscillation] are typically needed to see critical behavior.
Further, to get good ensemble averages, it seems that $\mathcal{O}(10)$ realizations of the initial
condition are needed for each of $\mathcal{O}(10)$ to $\mathcal{O}(100)$ realizations of natural periods.
In other words, $\mathcal{O}(100)$ to $\mathcal{O}(1000)$ realizations are needed for
lattice sizes and integration times of the above order og magnitude.
This is probably the reason why almost no clear-cut numerical results regarding
the existence or non-existence of phase transitions in oscillator lattices
have been presented in the past (section \ref{latticeintro}).
The data presented in this paper should be seen as an initial overview
of the behavior of some relevant quantities. A more detailed
study of each quantity is needed. In particular, the number of realizations of natural periods
has to be increased.

To put the theory to further test, it goes without saying that simulations
of oscillator lattices with dimensions other than $D=2$ are called for.
Perhaps the quantities used in this paper can be used to find an
answer to the long standing question whether there is a transition to
macroscopic synchronization in the three-dimensional Kuramoto model.

It is worth noting that the relevance of the
second critical coupling $g_{c2}$ is established in this study. It was first described in Ref. \cite{ptorus},
but there a density function $\mathcal{D}_{\omega}$ of natural frequencies with finite support was used.
Here, I find that it is present even if $\mathcal{D}_{\omega}$ has tails.
It is therefore a more generic transition than that to $R=1$ in the globally
coupled Kuramoto or Winfree models \cite{aria}, appearing when $\mathcal{D}_{\omega}$ has no tails.
The nature of the transition at $g_{c2}$ is a subject for future work,
and the question whether there is an additional non-trivial fixed point associated
with this transition is left unanswered.

Theory and numerics taken together indicate that the renormalization flow goes
towards increasing $g$ for $g>g_{c1}$ (Fig. \ref{Rfig20}). I judge that both
phase 2 and phase 3 are supercritical, and in section \ref{rflow}
I give a technical argument why this is so.
Here, a qualitative argument is presented why an oscillator lattice cannot be subcritical above $g_{c1}$,
that is, why correlation functions cannot have exponential tails, corresponding to finite correlation lenghts.

Correlation lengths relate to the typical distance a perturbation or fluctuation spreads.
Let us compare with the Ising model,
which is subcritical both above and below the critical temperature $T_{c}$.
In the ordered phase below $T_{c}$, most spins are aligned. Let us introduce a perturbation in the form of
a spin with opposite direction. Such a spin increases the probability that a neighbor
spin will also flip. The perturbation tends to spread. However, the lower the temperature,
the smaller the probability that the neighbor will flip, according to the Boltzmann
distribution. Thus, a typical perturbation spreads
shorter distances, and the correlation length drops.

The situation is quite different
in the ordered phase of an oscillator lattice, where I am thinking mainly of states
with partial frequency synchronization ($0<r<1$). A perturbation in such a lattice corresponds to an
oscillator $k$ that runs at a different frequency. This perturbation spreads to the rest
of the lattice via the coupling functions, which will not vary
with the entrained frequency. Assume for simplicity that the coupling has the form
$g \varphi_{lk}(\phi_{l}-\phi_{k})$. The peak magnitude of this perturbation can only increase with $g$,
since the
argument takes on all values in the range $[0,1)$ because $\Omega_{l}$ and $\Omega_{k}$
are assumed to be different. Thus, if the correlation lengths are infinite at a critical
coupling $g_{c1}$, they should stay infinite even if $g>g_{c1}$.

In conclusion, I hope that this study will inspire further theoretical and numerical work
on macroscopic synchronization in oscillator lattices. Unfortunately, the understanding of
these systems has fallen way behind the understanding of globally coupled oscillator networks.
A better understanding of oscillator lattices should also promote the
understanding of transitions to macroscopic synchronization in complex networks,
since the topology of these often can be seen as lying in between the topologies of
the lattice and the fully connected network.

This study was supported in part by the Royal Swedish Academy of Sciences
(Stiftelsen G. S. Magnusons fond).


\begin{thebibliography}{10}
\bibitem{aboutmean}Numerical evidence in Ref. \cite{ptorus} indicates that
frequency clusters are not completely stable in the parameter region with $0<r<1$ (phase 2),
even after the intial transient. One possibilty is that $\Omega_{k}$ does not exist
in this region, another is that it is identical for all oscillators. In the region with
$r=0$ (phase 1), it seems that the clusters are stable, and that one may
identify $\Omega_{k}=\lim_{t\rightarrow\infty}(d\phi_{k}/dt)$. This identification is
implicitly made in some derived formulae. If it is not true,
$\Omega_{k}$ should be replaced by $d\phi_{k}/dt$ in these expressions (for arbitrary $t$).
In that case I also judge that the numerical estimations of $\Omega_{k}$ [mean frequency during time $1000$
after a transient of $T_{tr}=\mathcal{O}(10^{5}$)] behaves in a way similar to $(d\phi_{k}/dt)(T_{tr})$,
since the instabilities of the clusters mainly appear on a larger time scale.
\bibitem{finiteg}If we let $g\rightarrow\infty$, then almost all $h_{k}\rightarrow\infty$
and $\mathrm{Var}[h_{k}]$ will not exist. Those $h_{k}=g\sum_{l\in n_{k}}f_{lk}(\phi_{l},\phi_{k})$
that might still be finite are those for which the phases in the $f_{lk}$s are fine tuned so
that the sum vanishes.
\end{thebibliography}
\end{document}